\documentclass[12pt]{iopart}     

\usepackage{iopams}

\def\tr{\tilde r}

\begin{document} 

\title{Solvable model for pair excitation in trapped Boson gas at zero temperature} 

\author{D Margetis}

\address{Department of Mathematics, and Institute for Physical Science and Technology,\\
University of Maryland, College Park, Maryland 20742, USA}

\ead{\mailto{dio@math.umd.edu}} 

\begin{abstract}
In Bose-Einstein condensation, a macroscopically large number of particles occupy
the same single-particle quantum state, $\Phi$. Our goal is to study time-dependent aspects of particle excitations 
to states other than $\Phi$ in trapped dilute atomic gases. 
We adopt the view that atoms are excited in pairs so that their scattering from $\Phi$ to positions 
${\bf x}$ and ${\bf y}$ at time $t$ is described by the pair-excitation function, $K_0({\bf x}, {\bf y},t)$.
We solve a nonlocal equation for $K_0({\bf x},{\bf y},t)$ under a slowly varying external potential 
by assuming that $\Phi({\bf x},t)$ satisfies a time-independent nonlinear Schr\"odinger equation. 
For zero initial excitation ($K_0\equiv 0$ at $t=0$) and sufficiently large $t$, we evaluate asymptotically
$K_0$ in terms of the one-variable Lommel function for any distance $|{\bf x}-{\bf y}|$. 
\looseness=-1
\end{abstract}

\vspace{2pc}

\submitto{\it J.~Phys.~A: Math.\ Theor.}
\maketitle 

\section{Introduction}

In Bose-Einstein condensation, atoms with integer spin (`Bosons') occupy a single-particle quantum state macroscopically.
This phenomenon, predicted by Bose~\cite{bose} and Einstein~\cite{einstein} for non-interacting particles over 80 years ago,
was observed experimentally in trapped dilute atomic gases in 1995~\cite{andersonetal95,davisetal95}.
Many similar experiments have followed~\cite{ketterleetal99}. These observations have renewed 
theoretical interest
in the Bose-Einstein condensation of systems that lack translational invariance. For recent reviews, see 
e.g.~\cite{pethick01,pitaevskii03,lieb05}. \looseness=-1

Theoretical studies of Bose-Einstein condensation for zero temperature often, though
by no means always~\cite{wu61,wu98,esry}, make use of a macroscopic wavefunction $\Phi({\bf x},t)$ that satisfies
a cubic nonlinear Schr\"odinger equation~\cite{wu61,wu98,gross,pitaevskii61}. Most recently, 
a mathematically rigorous derivation of this equation was given in the limit of an infinitely large number
of interacting particles~\cite{yau}.\looseness=-1

The use of the nonlinear Schr\"odinger equation for $\Phi({\bf x},t)$ has
been deemed adequate for many experimental situations at extremely low temperatures~\cite{ketterleetal99}. 
However, this description is fundamentally incomplete even at zero temperature 
for a finite number of interacting particles. In principle,
atoms are excited from $\Phi$ to other states. This many-body process is not accounted for by
the nonlinear Schr\"odinger equation. \looseness=-1

Particle excitations in Bose-Einstein condensation were described 
systematically by Lee, Huang and Yang~\cite{leehuangyang} for systems with translational invariance 
and periodic boundary conditions. In this setting, where there is no external potential,
atoms are primarily excited {\it in pairs} from the (lowest) 
state of zero momentum to states with opposite momenta~\cite{leehuangyang}.
This process leads to phonons and sound vibrations~\cite{leehuangyang}. \looseness=-1

Wu~\cite{wu61} extended the theory of~\cite{leehuangyang} to systems
that lack translational symmetry. A key ingredient of his formulation is the pair-excitation 
function, $K_0({\bf x},{\bf y},t)$, which
describes the scattering of atoms in pairs from $\Phi$
to other states at positions ${\bf x}$ and ${\bf y}$; see~(\ref{eq:Psi-Ans}) below.
This formulation yields coupled nonlocal equations for $\Phi({\bf x},t)$
and $K_0({\bf x},{\bf y},t)$~\cite{wu61,wu98}, and therefore transcends other treatments
based solely on the nonlinear Schr\"odinger equation. 
However, solutions of the resulting coupled equations have
remained largely elusive. By approximately decoupling the two equations, 
Wu described a time-independent solution for $K_0$ in a slowly
varying trapping potential~\cite{wu98}.\looseness=-1

In this paper we address analytically aspects of the {\it time dependence} of $K_0$ 
for a system of trapped interacting atoms at zero temperature. 
Our starting point is the coupled system of $\Phi$ and $K_0$~\cite{wu61,wu98},
which we simplify by assuming that $\Phi$ satisfies the nonlinear Schr\"odinger equation
independently. The resulting nonlocal
equation for $K_0$ is solved approximately for a time-independent, slowly varying external potential.
In this context, $\Phi$ is treated as a given, variable coefficient of the equation for $K_0$.
We carry out a large-$t$ asymptotic analysis that shows how $K_0$ approaches the steady state
if $K_0\equiv 0$ at $t=0$, i.e.,
initially all atoms occupy the single-particle state $\Phi$.\looseness=-1 

Our study of pair excitation is motivated by three broader questions.
The first question concerns the precise manipulation of atomic gases at very low temperatures. 
In current experimental setups, a rich variety of effects are observed 
including the depletion of the macroscopic state $\Phi$; see e.g.~\cite{xuetal06}. 
Therefore, it is of interest to refine our understanding of how atomic interactions influence
macroscopic properties of the Boson gas. The second question concerns the possible extension of
the concept of a phonon to systems that lack translational symmetry.
The third question concerns the analysis of coupled nonlocal equations. Pair excitation leads to integrodifferential equations 
of motion that are as yet unexplored. Here, we seek particular solutions but abandon
mathematical rigor. Analytical solutions of simplified models such as the one in this paper 
may be a guide for future rigorous studies of similar models.\looseness=-1 

The starting point is Wu's formulation~\cite{wu61,wu98} for a system of $N$ pairwise
interacting Bosons at positions $\{{\bf x}_i\}$. By units with $\hbar=2m=1$
($\hbar$: Planck's constant, $m$: atomic mass), the many-body Hamiltonian reads
\begin{equation}
H=\sum_{i=1}^N[-\Delta_i+V_e({\bf x}_i)]+4\pi a\sum_{i\neq j}\delta({\bf x}_i-{\bf x}_j)\,
\frac{\partial}{\partial x_{ij}}x_{ij}~,\quad
a>0~,
\label{eq:Hamiltonian}
\end{equation}
where $\Delta_i$ is the Laplacian corresponding to ${\bf x}_i$,
$a$ is the scattering length, $V_e$ is the external potential, and $x_{ij}:=|{\bf x}_i-{\bf x}_j|$.
By use of quantized fields~\cite{berezin}, the $N$-body wavefunction of the Boson system is assumed to be of the form~\cite{wu61,wu98}
\begin{equation}
\Psi(t)= {\mathcal N} (t)\,e^{{\mathcal P}
(t)}\{(N!)^{-1/2}a^*_0(t)^N|{\rm vac}\rangle\}~.
\label{eq:Psi-Ans}
\end{equation}
In this equation, ${\mathcal N}$ is a normalization constant, $a_0^*(t)$ is the creation operator for
the state $\Phi$, $|{\rm vac}\rangle$ is the vacuum state, and the operator ${\mathcal P}(t)$ is
\begin{equation}
{\mathcal P} (t)\propto \int\!\int {\rm d}{\bf x}\,{\rm d}{\bf y}\,
\psi^*_1({\bf x},t)\psi^*_1({\bf y},t)\,K_0({\bf x},{\bf y},t)
a_0(t)^2~.\label{eq:pair-def}
\end{equation}Here, $\psi_1^*({\bf x},t)$ is the creation field operator corresponding
to the space orthogonal to $\Phi$. Notice that the integrand of ${\mathcal P}$ in~(\ref{eq:pair-def}) describes
the annihilation of two particles from $\Phi$ with the simultaneous creation of two particles
at positions ${\bf x}$ and ${\bf y}$ at other states. If ${\mathcal P}(t)\equiv 0$ (or, $K_0\equiv 0$)
then no atoms are excited from $\Phi$. So, the many-body wavefunction reduces to a tensor product
of single-particle states each of which is $\Phi$. \looseness=-1

The functions $\Phi({\bf x},t)$ and $K_0({\bf x},{\bf y},t)$ are found to satisfy
two coupled nonlinear equations~\cite{wu61,wu98}. These equations take the form~\cite{wu61,wu98}
\begin{eqnarray}
i\,\partial_t\Phi=[-\Delta_x+V_e({\bf x}) 
+8\pi a\rho_0|\Phi|^2-4\pi a\rho_0\zeta(t)]\Phi
+N^{-1}{\mathcal W}_1({\bf x},t),\label{eq:Phi}
\end{eqnarray}
\begin{eqnarray}
\lefteqn{[i\partial_t-2E(t)]K_0=-(\Delta_x +\Delta_y) K_0 +8\pi a\rho_0\,\Phi({\bf x},t)^2\delta({\bf x}-{\bf y})}
\nonumber\\
&&\mbox{}+\{-Z(t)+V_e({\bf x})+V_e({\bf y})+16\pi a\rho_0
[|\Phi({\bf x},t)|^2+|\Phi({\bf y},t)|^2]\}K_0\nonumber\\
&&\mbox{}+8\pi a\rho_0\int {\rm d}{\bf z}\,\Phi^*({\bf z},t)^2
K_0({\bf x},{\bf z},t)K_0({\bf y},{\bf z},t)+N^{-1}{\mathcal W}_2({\bf x},{\bf y},t),\label{eq:K}
\end{eqnarray}
where $\rho_0$ is the (constant) equilibrium density of the system and $i^2=-1$.
The functions $\zeta(t)$, $Z(t)$ and $E(t)$
are integrals of $\Phi$ to be defined in section~\ref{sec:formulation}, and
${\mathcal W}_{l}$ ($l=1,\,2$) are nonlinear functionals of $\Phi$ and $K_0$. We note in passing that
$\rho_0$ can be eliminated from the equations of motion by using the variable 
$\tilde\Phi:=\rho_0^{1/2}\Phi$ in place of $\Phi$~\cite{wu98}.
The system of~(\ref{eq:Phi}) and~(\ref{eq:K}) form a nontrivial extension of the cubic nonlinear Schr\"odinger equation
for $\Phi$, which results from~(\ref{eq:Phi}) by setting ${\mathcal W}_1\equiv 0$.

In this paper we study~(\ref{eq:K}) when $V_e({\bf x})$ is sufficiently slowly varying
by enforcing ${\mathcal W}_1\equiv {\mathcal W}_2\equiv 0$ and 
$\Phi({\bf x},t)=e^{-iEt}\Phi({\bf x})$, where $E$ is the energy per particle of the 
macroscopic state. Despite these crucial
simplifications, (\ref{eq:K}) remains nonlocal and coupled to the solution
of the nonlinear Schr\"odinger equation. Equation
(\ref{eq:K}) can be solved exactly in terms of a spherically symmetric Fourier integral. We evaluate this
integral for sufficiently large $t$ for any value of $r=|{\bf x}-{\bf y}|$
by expanding the integrand in an appropriate series and applying the standard stationary-phase method~\cite{cheng}. 
As $t$ and $r$ vary, significant contributions to $K_0$ occur when a stationary-phase point
coalesces with the origin in Fourier space, which is the endpoint of integration.
It is of some interest to recognize this coalescence as an `elementary catastrophe',
by analogy with studies of diffraction phenomena by Berry~\cite{berry}. The time-dependent 
pair-excitation function is described by a solution of Lommel's 
differential equation~\cite{chandr}.\looseness=-1

We stress that our work relies directly on ansatz~(\ref{eq:Psi-Ans})
for the many-body wavefunction: the pair-excitation function, $K_0({\bf x},{\bf y},t)$, is used as
a variable in addition to the macroscopic wavefunction, $\Phi({\bf x},t)$. 
This approach is distinctly different
from methods based on the hydrodynamic theory of superfluids, e.g.~\cite{nozieres,pitaevskii98}.
Once $K_0$ and $\Phi$ are known, many-body properties of the Boson gas
can in principle be computed via~(\ref{eq:Psi-Ans}).\looseness=-1

A few remarks on our main simplifying assumptions are in order. 
(i) $K_0$ does not act back on $\Phi$ since we take $\mathcal W_1\equiv 0$ in~(\ref{eq:Phi}). 
Thus, the issue of how $\Phi$ is 
modified by $K_0$ is left unresolved here. Decoupling $\Phi$
from $K_0$ in this sense may pose a limitation on the time scale for the validity of our results
in a physical setting. 
Therefore, the large-$t$ limit studied here should be interpreted with caution, especially
in connection to experiments.
(ii) The external potential, $V_e$, is time independent and slowly varying, 
and $\Phi$ is taken to be time independent. A reasonable approximation for $\Phi$ then results by neglecting  
the Laplacian in the nonlinear Schr\"odinger equation~\cite{wu98}. This simplification, sometimes referred to
as the `Thomas-Fermi approximation' in the context of Bose-Einstein 
condensation~\cite{pethick01,pitaevskii03}, amounts
to seeking an outer solution in the sense of singular perturbation~\cite{cheng}. 
The effect on the motion of $K_0$
of possible `boundary layers' for $\Phi$, where the Laplacian needs to be retained~\cite{wu98}, 
is not addressed by our analysis.\looseness=-1

The remainder of the paper is organized as follows. 
In section~\ref{sec:formulation} we review the simplified equations of motion.
In section~\ref{sec:trans-inv} we focus on the instructive case with translational invariance where
$V_e$ and $\Phi$ are constants: in section~\ref{subsec:exact} we derive an integral
formula for $K_0({\bf x},{\bf y},t)$; and in section~\ref{subsec:asymp} we evaluate the
requisite integral for large $t$. 
In section~\ref{sec:slow} we focus on the case with a slowly varying external potential:
in section~\ref{subsec:phi-app} we describe an approximate solution for $\Phi({\bf x},t)$ revisiting~\cite{wu98};
and in section~\ref{subsec:K-app} we show that the time-dependent $K_0$ can be determined with minor 
modifications of the analysis of section~\ref{sec:trans-inv}. In section~\ref{sec:conclusion}
we summarize our results and discuss related open questions. The units with $\hbar=2m=1$
are used throughout the analysis.

\section{Background theory: simplified equations of motion}
\label{sec:formulation}

In this section we review the simplified equations of motion for $\Phi$ and $K_0$.
 which are derived in~\cite{wu98}. These
equations form the starting point of the analysis in sections~\ref{sec:trans-inv} and~\ref{sec:slow}.

The macroscopic wave function $\Phi({\bf x},t)$ satisfies the nonlinear Schr\"odinger equation
\begin{equation}
i\partial_t\Phi({\bf x},t)=[-\Delta_{\bf x}+V_e({\bf x}) 
+8\pi a\rho_0|\Phi({\bf x},t)|^2-4\pi a\rho_0\zeta(t)]\Phi({\bf r},t)~,\label{eq:Phi-NLSE}
\end{equation}
where $\rho_0=N/\Omega$, $\Omega$ is the volume of the system, 
\begin{equation}
\zeta(t):=\Omega^{-1}\int{\rm d}{\bf x}\,|\Phi({\bf x},t)|^4~,
\label{eq:zeta-def}
\end{equation}
and $\Phi$ is subject to the normalization condition
\begin{equation}
\Omega^{-1}\int{\rm d}{\bf x}\,|\Phi({\bf x},t)|^2=1~.
\label{eq:Phi-norm}
\end{equation}

The pair-excitation function $K_0({\bf x},{\bf y},t)$ solves the 
integrodifferential equation~\cite{wu98} 
\begin{eqnarray}
\lefteqn{[i\partial_t-2 E(t)]K_0({\bf x},{\bf y},t)=-(\Delta_x+\Delta_y)K_0({\bf x},{\bf y},t)
+8\pi a\rho_0\Phi({\bf x},t)^2}
\nonumber\\
&&\mbox{}\times\delta({\bf x}-{\bf
y})+\biggl\{-2\bar\zeta(t)-16\pi a\rho_0\,\zeta(t)-2\zeta_e(t)+V_e({\bf x})+V_e({\bf y})
\nonumber\\ 
&&\mbox{} +16\pi
a\rho_0\,[|\Phi({\bf x},t)|^2+|\Phi({\bf y},t)|^2]\biggr\}K_0({\bf x},{\bf y},t)+8\pi a\rho_0
\nonumber\\ 
&&\mbox{} \times \int {\rm d}{\bf z}\,
\Phi^*({\bf z},t)^2 K_0({\bf x},{\bf z},t) K_0({\bf y},{\bf z},t)
-8\pi a\rho_0\Omega^{-1}\biggl\{ \Phi({\bf x},t)\Phi({\bf y},t)
\nonumber\\
&&\mbox{} \times[|\Phi({\bf x},t)|^2+|\Phi({\bf y},t)|^2
-\zeta(t)]+\Phi({\bf x},t)\int{\rm d}{\bf z}\,K_0({\bf y},{\bf z},t)|\Phi({\bf z},t)|^2
\nonumber\\
&&\mbox{} \times\Phi^*({\bf z},t)+\Phi({\bf y},t)\int{\rm d}{\bf z}\,K_0({\bf x},{\bf z}, t)|\Phi({\bf z},t)|^2
\Phi^*({\bf z},t)\biggr\}~,
\label{eq:K0-IDE}
\end{eqnarray}
where 
\begin{equation}
E(t):=\Omega^{-1}\int {\rm d}{\bf x}\,\Phi^*({\bf x},t)\ i\partial_t\Phi({\bf x},t)~,\label{eq:Et-def}
\end{equation}
\begin{equation}
\bar\zeta(t):=\Omega^{-1}\int{\rm d}{\bf x}\,|\nabla\Phi({\bf x},t)|^2~,\label{eq:barz-def}
\end{equation}
\begin{equation}
\zeta_e(t):=\Omega^{-1}\int{\rm d}{\bf x}\,V_e({\bf x})|\Phi({\bf x},t)|^2~.
\label{eq:zetae-def}
\end{equation}

For sufficiently large $N$ and fixed $\rho_0$, (\ref{eq:K0-IDE}) is further simplified:
the term proportional to $\Omega^{-1}$ in the right-hand side is neglected.
Thus,~(\ref{eq:K0-IDE}) reduces to the equation~\cite{wu98}
\begin{eqnarray}
\lefteqn{[i\,\partial_t-2E(t)]K_0({\bf x},{\bf y},t)=-(\Delta_x
+\Delta_y)K_0({\bf x},{\bf y},t)
+8\pi a\rho_0\Phi({\bf x},t)^2}
\nonumber\\
&&\mbox{}\times\delta({\bf x}-{\bf
y})+\biggl\{-2\bar\zeta(t)-16\pi a\rho_0\,\zeta(t)-2\zeta_e(t)+V_e({\bf x})+V_e({\bf y})
\nonumber\\ 
&&\mbox{} +16\pi
a\rho_0\,[|\Phi({\bf x},t)|^2+|\Phi({\bf y},t)|^2]\biggr\}K_0({\bf x},{\bf y},t)
\nonumber\\ 
&&\mbox{} +8\pi a\rho_0 \int {\rm d}{\bf z}\,
\Phi^*({\bf z},t)^2 K_0({\bf x},{\bf z},t) K_0({\bf y},{\bf z},t)~.
\label{eq:K0-IDE-app}
\end{eqnarray}

Following~\cite{wu98} we use the center-of-mass coordinates,
\begin{equation}
{\bf r}:={\bf x}-{\bf y}~,\quad {\bf R}:=\frac{{\bf x}+{\bf y}}{2}~,
\label{eq:com}
\end{equation}
and set
\begin{equation}
K_0({\bf x},{\bf y},t)=:\mathcal K({\bf R},{\bf r},t)~.
\label{eq:calK0}
\end{equation}
Hence,~(\ref{eq:K0-IDE-app}) is recast to the equation
\begin{eqnarray}
\lefteqn{[i\,\partial_t-2E(t)]{\mathcal K}({\bf R},{\bf r},t)=-\big(\textstyle{\frac{1}{2}}\Delta_R
+2\Delta_r\big){\mathcal K}+8\pi a\rho_0\Phi({\bf R},t)^2\,\delta({\bf r})}\nonumber\\
&& +\big\{-2\bar\zeta(t)-16\pi a\rho_0\zeta(t)-2\zeta_e(t)+V_e\big({\bf R}+\textstyle{\frac{1}{2}}{\bf r}\big)+V_e\big({\bf R}
-\textstyle{\frac{1}{2}}{\bf r}\big)\nonumber\\
&& +16\pi a\rho_0\big[|\Phi\big({\bf R}+\textstyle{\frac{1}{2}}{\bf r},t\big)|^2+|\Phi\big({\bf R}-\textstyle{\frac{1}{2}}{\bf r},t\big)|^2\big]\big\}
{\mathcal K}({\bf R},{\bf r},t)\nonumber\\
&& +8\pi a\rho_0 \int {\rm d}{\bf w}\,\Phi^*\big({\bf R}-\textstyle{\frac{1}{2}}{\bf r}-{\bf w},t\big)^2
{\mathcal K}({\bf R}+\textstyle{\frac{1}{2}}{\bf r}-\textstyle{\frac{1}{2}}{\bf w},{\bf w},t)\nonumber\\
\mbox{} && \hskip40pt \times{\mathcal K}({\bf R}-\textstyle{\frac{1}{2}}{\bf w},{\bf w}-{\bf r},t)~.
\label{eq:K0-com}
\end{eqnarray}

In the next section, we focus on solving~(\ref{eq:Phi-NLSE}) and~(\ref{eq:K0-com}).

\section{Translational invariance}
\label{sec:trans-inv}

In this section we enforce translational invariance and periodic boundary conditions
taking $V_e={\rm const}$. First, we set $\Phi={\rm const.}$ and study the resulting equation
of motion for $K_0$. By invoking the Fourier transform of $K_0({\bf x},{\bf y},t)$ in ${\bf x}-{\bf y}$ we find an integral 
representation for $K_0$. Second, we evaluate this integral for sufficiently large $t$.
The results obtained here are useful in section~\ref{sec:slow} where $V_e({\bf x})$
is taken to be slowly varying.

\subsection{Solution for pair excitation}
\label{subsec:exact}

In view of condition~(\ref{eq:Phi-norm}) we set
\begin{equation}
\Phi({\bf x},t)\equiv 1\qquad t\ge 0~.\label{eq:Phi_eq_1}
\end{equation}
By (\ref{eq:zeta-def}), (\ref{eq:Et-def}) and~(\ref{eq:barz-def}), we have
\begin{equation}
\zeta(t)=1~,\quad E(t)=0~,\quad \bar\zeta(t)= 0~.\label{eq:E-zeta-trans}
\end{equation}
Thus,~(\ref{eq:Phi-NLSE}) along with~(\ref{eq:zetae-def}) entail 
\begin{equation}
V_e=-4\pi a\rho_0\equiv\zeta_e(t)~.\label{eq:Ve-trans}
\end{equation}
The goal is to solve~(\ref{eq:K0-IDE-app}) under~(\ref{eq:Phi_eq_1})--(\ref{eq:Ve-trans}).

By the initial condition
\begin{equation}
\mathcal K({\bf R},{\bf r},0)=f({\bf r})~,\label{eq:init}
\end{equation}
$\mathcal K({\bf R},{\bf r},t)$ should be ${\bf R}$-independent for $t>0$.
With the replacement $\mathcal K({\bf R},{\bf r},t)=\mathcal K({\bf r},t)$, 
(\ref{eq:K0-com}) becomes
\begin{eqnarray}
\lefteqn{i\,\partial_t{\mathcal K}({\bf r},t)=-2\Delta_r{\mathcal K}({\bf r},t)+8\pi a\rho_0\,\delta({\bf r})
+16\pi a\rho_0 {\mathcal K}({\bf r},t)}\nonumber\\
&&+8\pi a\rho_0\int {\rm d}{\bf w}\ {\mathcal K}({\bf w})\,{\mathcal K}({\bf w}-{\bf r},t)~.
\label{eq:K0-com-sph}
\end{eqnarray}
In addition, we have the symmetry relation $\mathcal K({\bf r},t)=\mathcal K(-{\bf r},t)$
assuming that $f({\bf r})=f(-{\bf r})$. Thus,
a convolution integral appears on the right-hand side of~(\ref{eq:K0-com-sph}).

Accordingly, we introduce the Fourier transform of $\mathcal K$ by
\begin{equation}
\widehat {\mathcal K}({\bf k},t)=\int {\rm d}{\bf r}\ \mathcal K({\bf r},t) \,e^{-i{\bf k}\cdot {\bf r}}~,
\label{eq:FT-K0}
\end{equation}
whence
\begin{equation}
{\mathcal K}({\bf r},t)=\int\frac{{\rm d}{\bf k}}{(2\pi)^3}\ e^{i{\bf k}\cdot {\bf r}}\, \widehat{\mathcal K}({\bf k},t)~.
\label{eq:Inv-FT}
\end{equation}
Hence,~(\ref{eq:K0-com-sph}) is transformed to
\begin{equation}
i\,\partial_t\widehat {\mathcal K}({\bf k},t)=8\pi a\rho_0 \widehat{\mathcal K}^2+2(k^2+8\pi a\rho_0)
\widehat {\mathcal K}+8\pi a\rho_0\quad (k=|{\bf k}|)~.
\label{eq:K0-FT-ODE}
\end{equation}

The steady state for~(\ref{eq:K0-com-sph}) stems from taking $\partial_t\widehat {\mathcal K}\equiv 0$
in~(\ref{eq:K0-FT-ODE}).
We interpret the resulting time independence as the long-time ($t\to +\infty$) limit, in an appropriate sense, 
of the solution to~(\ref{eq:K0-FT-ODE}). 
The corresponding limit for $\widehat {\mathcal K}$ reads
\begin{equation}
\widehat K({\bf k},+\infty)=-\frac{ k^2+8\pi a\rho_0-k\sqrt{k^2+16\pi a\rho_0}\,}{8\pi a\rho_0}=:\widehat g_0(k).
\label{eq:steady-k}
\end{equation} 
This time-independent solution is derived in~\cite{wu98} for a slowly-varying
external potential. The inversion of this $\widehat g_0(k)$ gives~\cite{wu98}
\begin{equation}
g_0(r)=\pi^{-2}(4\pi a\rho_0)^{3/2}\,\chi(r)^{-1}\ {\rm Im}\big[S_{0,4}
\big(i\chi(r)\big)-S_{0,0}(i\chi(r))\big],
\label{eq:steady-r}
\end{equation}
where $g_0(r):=\mathcal K({\bf r},+\infty)$, $r=|{\bf r}|$, $S_{\mu,\nu}$ denotes the Lommel function~\cite{bateman}, and
\begin{equation}
\chi(r):=(16\pi a\rho_0)^{1/2}\,r~.\label{eq:chi-def}
\end{equation}

We turn our attention to the time-dependent $\mathcal K$.
Equation~(\ref{eq:K0-FT-ODE}) can be solved explicitly.
Details of the integration procedure are given in~\ref{app:A}.
In view of~(\ref{eq:init}), the solution reads
\begin{equation}
\widehat {\mathcal K}({\bf k},t)=\widehat{g}_0(k)-[1-\widehat{g}_0(k)^2]\,
\frac{p({\bf k})\,e^{-2i\omega(k)\,t}}{1-
\widehat{g}_0(k)\,p({\bf k})e^{-2i\omega(k)\,t}}~,
\label{eq:hatK0-soln}
\end{equation}
where
\begin{equation}
\omega(k):=k\sqrt{k^2+16\pi a\rho_0}=
[\widehat g_0(k)-\widehat g_0(k)^{-1}]~,\label{eq:omega-def}
\end{equation}
\begin{equation}
p({\bf k}):=\frac{\widehat g_0(k)-\widehat f({\bf k})}{1-\widehat g_0(k)\widehat f({\bf k})}~,
\label{eq:p-def}
\end{equation}
and $\widehat f({\bf k})$ is the Fourier transform of the initial data, equation~(\ref{eq:init}).
Note that $\,-1\le\widehat g_0<0$ if $k\ge 0$. The function $\omega(k)$ gives the
phonon spectrum, or the energy of excitation from zero momentum into a state of momentum ${\bf k}$~\cite{leehuangyang}.

We next apply spherically symmetric initial data, $f({\bf r})\equiv f_0(r)$ in~(\ref{eq:init}).
Thus, $p({\bf k})$ is replaced by the function
\begin{equation}
p({\bf k})=p_0(k):=\frac{\widehat g_0(k)-\widehat f_0(k)}{1-\widehat g_0(k)\widehat f_0(k)}~,
\label{eq:p0}
\end{equation}
where $\widehat f_0(k)$ is the Fourier transform of $f_0(r)$. Equation~(\ref{eq:hatK0-soln}) becomes
\begin{equation}
\widehat {\mathcal K}({\bf k},t)=\widehat{g}_0(k)-[1-\widehat{g}_0(k)^2]\,
\frac{p_0(k)\,e^{-2i\omega(k)\,t}}{1-
\widehat{g}_0(k)\,p_0(k)e^{-2i\omega(k)\,t}}~.
\label{eq:hatK0-symm}
\end{equation}

We note in passing that the right-hand side of~(\ref{eq:hatK0-symm})
can be continued analytically in the complex $k$-plane where singularities are located only
in the first and third quadrants. We refer the reader to~\ref{app:B} for details when $f\equiv 0$. 
This observation is relevant to the asymptotic calculation of section~\ref{subsec:asymp} and the
long-time limit~(\ref{eq:steady-k}): by deforming the inversion path for $\widehat{\mathcal K}$
slightly below the positive real $k$-axis, we can expicitly ensure that $e^{-2i\omega(k) t}\to 0$ as $t\to +\infty$. 
Consistent with definition~(\ref{eq:FT-K0}), 
this limit can be generalized to include solution~(\ref{eq:hatK0-soln}) for a reasonably 
wide class of initial data $\widehat f({\bf k})$. 
The limit (\ref{eq:steady-k}) is then readily recovered.

The inversion of $\hat\mathcal K({\bf k},t)$ is expressed by the integral
\begin{eqnarray}
\mathcal K({\bf r},t)&=&\frac{1}{2\pi^2 r}
\int\limits_0^{+\infty}{\rm d}k\ k\sin(kr)\nonumber\\
&&\mbox{}\times\biggl\{\widehat g_0(k)-[1-\widehat g_0(k)^2]\,
\frac{p_0(k)\,e^{-2i\omega(k)\,t}}{1-
\widehat g_0(k)\,p_0(k)e^{-2i\omega(k)\,t}}\biggr\}~.\label{eq:K0-exact}
\end{eqnarray} 
We have not been able to compute this integral exactly in simple closed form in terms
of known special functions. 

Equation (\ref{eq:K0-exact}) is written in the convenient form\looseness=-1
\begin{eqnarray}
\lefteqn{\Lambda(r,t):=(16\pi a\rho_0)^{-3/2}
[\mathcal K({\bf r},t)-\mathcal K({\bf r},+\infty)]=-\frac{(16\pi a\rho_0)^{-3/2}}{2\pi^2 r}}\nonumber\\
&&\mbox{}\times\int\limits_0^{+\infty}{\rm d}k\ k\sin(k r)
[1-\widehat g_0(k)^2]\,\frac{p_0(k)\,e^{-2i\omega(k)\,t}}{1-
\widehat g_0(k)\,p_0(k)e^{-2i\omega(k)\,t}}~,\label{eq:Lambda-def}
\end{eqnarray}where $\Lambda$ is non-dimensional and 
$\mathcal K({\bf r},+\infty)=g_0(r)$ is defined by~(\ref{eq:steady-r}).
The function $\Lambda(r,t)$ expresses the deviation of the pair excitation function, $\mathcal K$,
from the steady-state solution, $g_0(r)$.

In the following, we focus on zero initial data, i.e. $f_0(r)\equiv 0$, by which all particles initially
occupy the state $\Phi$; cf~(\ref{eq:Psi-Ans}). Hence, we have
\begin{equation}
p_0(k)\equiv \widehat g_0(k)~.\label{eq:simpl-p}
\end{equation}
By the change of variable $k=(16\pi a\rho_0)^{1/2}\sinh\eta$,~(\ref{eq:Lambda-def}) becomes
\begin{equation}
\Lambda(r,t)=\frac{1}{2\pi^2 \tr}\int\limits_0^{+\infty}{\rm d}\eta\, [\sinh(2\eta)]^2\,\sin(\tr\sinh\eta)\,
\frac{e^{-4\eta}\,e^{-i\tau\sinh(2\eta)}}{1-e^{-4\eta-i\tau\sinh(2\eta)}},
\label{eq:Lambda-eta}
\end{equation}
where the non-dimensional variables $\tr$ and $\tau$ are defined by
\begin{equation}
\tr:=(16\pi a\rho_0)^{1/2}\,r~,\qquad \tau:=(16\pi a\rho_0)\,t~.
\label{eq:tau-r}
\end{equation}
As discussed in~\ref{app:B}, the path of integration in~(\ref{eq:Lambda-eta})
can be deformed slightly below the positive real axis since there are no singularities
in that part of the $\eta$-plane.

\subsection{Asymptotic analysis for $t\gg (16\pi a\rho_0)^{-1}$}
\label{subsec:asymp}

In this subsection, we determine $\Lambda(r,t)$ for zero initial data and any $r$ 
when $t\gg (16\pi a\rho_0)^{-1}$. 
In view of~(\ref{eq:Lambda-def}), a key idea is that, if $|p_0(k)|< 1$, we can expand
$[1-\widehat g_0(k)p_0(k)e^{-i2\omega(k) t}]^{-1}$ in a geometric series. Thus, we have
\begin{equation}
\widehat{\mathcal K}({\bf k},t)-{\widehat g}_0(k)=(4\pi a\rho_0)^{-1}\omega(k)
\sum_{l=1}^{+\infty}[{\widehat g}_0(k) p_0(k)]^l\,e^{-il\,2\omega(k)\,t}~,
\label{eq:hatK0-exp}
\end{equation}where we used~(\ref{eq:omega-def}). 
When both $t$ and $r$ are large enough, each corresponding integral under the summation sign for $\Lambda(r,t)$ 
can be evaluated by the method of stationary
phase~\cite{cheng}. As $t$ and $r$ vary, a critical point coalesces with the endpoint
of integration, $k=0$, giving rise to a behavior described by the one-variable 
Lommel function~\cite{bateman}.

We proceed to carry out this program by resorting
to~(\ref{eq:Lambda-eta}), using the non-dimensional variables $\tau$ and $\tr$ where $\tau\gg 1$. 
The dependence on $r$, $t$ and $a\rho_0$ can be restored
via definitions~(\ref{eq:tau-r}) once the calculation has been completed.

\subsubsection{Case $\tr=O(1)$.}
\label{sssec:rordt}

When $\tr$ is fixed, the major contribution to integration in
(\ref{eq:Lambda-eta}) comes from the vicinity of $\eta=0$. We obtain the expression
\begin{eqnarray}
\Lambda(r,t) &\sim&  \frac{1}{2 \pi^2 \tr}\int\limits_0^{+\infty}{\rm d}\eta\  
(2\eta)^2\,(\tr\eta)\,\frac{e^{-i\tau\,(2\eta)}}{1- e^{-i \tau (2\eta)}}\nonumber\\
{}&=& \frac{1}{8\pi^2}\,\frac{1}{\tau^4}\,
\psi'''(1)=\frac{\pi^2}{120}\,\frac{1}{\tau^4}~,\label{eq:Lambda-r-fix}
\end{eqnarray}
where $\psi(z)$ is the logarithmic derivative of the Gamma function 
and $\psi^{\prime\prime\prime}(1)=\pi^4/15$ is the third derivative of $\psi(z)$ at $z=1$~\cite{batemanI}. 
Formula~(\ref{eq:Lambda-r-fix}) is not particularly informative because of the absence
of $r$-dependence. 

It of some interest to re-derive~(\ref{eq:Lambda-r-fix}) in light of~(\ref{eq:hatK0-exp}).
By writing
\begin{eqnarray}
\big[1-e^{-4\eta}\,e^{-i \tau\sinh (2\eta)}\big]^{-1}&=&
\sum_{l=0}^{M-1}e^{-4l\eta}\,e^{-il  \tau\sinh (2\eta)}\nonumber\\
&&\mbox{}+ \frac{e^{-4M\eta}\,e^{-iM t\sinh (2\eta)}}{1-e^{-4\eta}\,e^{-i
\tau\sinh (2\eta)}}~,\quad M\ge 1~,\label{eq:exp-eta}
\end{eqnarray}
$\Lambda(r,t)$ from~(\ref{eq:Lambda-eta}) becomes
\begin{equation}
\Lambda(r,t)=\frac{1}{2\pi^2 \tr}\,\left[\sum_{l=1}^{M} I_l(\tr, \tau)
+R_{M}(\tr,\tau)\right]~,\label{eq:Lambda-Il} 
\end{equation}
where
\begin{equation}
I_l(\tr,\tau)=\int\limits_0^{+\infty}{\rm d}\eta\ [\sinh (2\eta)]^2\ \sin(\tr\sinh
\eta)\, e^{-4l\eta}\,e^{-il\, \tau\sinh (2\eta)}~,\label{eq:Il-def}
\end{equation}
\begin{equation}
R_{M}=\int\limits_0^{+\infty}{\rm d}\eta\,[\sinh (2\eta)]^2\sin(\tr\sinh\eta)\,
\frac{e^{-4(M+1)\eta}\,e^{-i(M+1) \tau\sinh (2\eta)}}{1-e^{-4\eta} \,e^{-i \tau\sinh
(2\eta)}}.\label{eq:RM-def}
\end{equation}

By deforming the integration path into the lower $\eta$-plane, we have~\cite{batemanI}
\begin{eqnarray}
R_{M}(\tr,\tau) &\sim&  4 r\int\limits_0^{\infty}{\rm d}y\ y^3\,\frac{e^{2i(M+1)\,(2y)}\,e^{-(M+1)
\tau\,(2y)}}{1-e^{- \tau\,(2y)}}\nonumber\\
{} &=& \frac{\tr}{4\tau^4} \,\psi'''\big((M+1)(1-2i/\tau)\big)
\sim \frac{\tr}{2 \tau^4}\,\frac{1}{(M+1)^3}~,\label{eq:RM-approx}
\end{eqnarray}
uniformly in $\tr$, where $M\gg 1$. The integral $I_l$ of~(\ref{eq:Il-def}) is
\begin{equation}
I_{l}(\tr,\tau)\sim \int\limits_0^{+\infty}{\rm d}\eta\ (2\eta)^2\ (\tr \eta)\,
e^{-il\, \tau\,(2\eta)}= \frac{3 \tr}{2l^4 \tau^4}~,\qquad \tr=O(1)~,\label{eq:Il-app}
\end{equation}
for any $l\ge 1$. So, by~(\ref{eq:Lambda-Il}) we have 
\begin{equation}
\Lambda(r,t)\sim\frac{1}{2\pi^2}\frac{3\tr}{2\tau^4}\left[\sum_{l=1}^M \frac{1}{l^4}+O\big(M^{-3}\big)\right]\to
\frac{\pi^2}{120}\,\tau^{-4}~,\quad M\to +\infty~.\label{eq:Lambda-rfix}
\end{equation}

In the following, we invoke~(\ref{eq:Lambda-Il}) and (\ref{eq:Il-def}) when 
$\tr$ is comparable to or larger than $\tau$.\looseness=-1

\subsubsection{Case $\tr\ge O(l\tau)$.}
\label{sssec:rgordt}

In this case we need to describe the $r$-dependence of $\Lambda$ by appropriately 
accounting for $\sin(\tr\sinh\eta)$ in~(\ref{eq:Lambda-eta}).
For this purpose, we apply~(\ref{eq:Lambda-Il}) for $M\to +\infty$:
\begin{equation}
\Lambda(r,t)=\frac{1}{2\pi^2  \tr}\sum_{l=1}^{\infty}I_l(\tr,\tau)~,\label{eq:Lambda-I}
\end{equation}
where
\begin{equation}
I_{l}(\tr,\tau)=I_{l,+}(\tr,\tau)-I_{l,-}(\tr,\tau)~,\label{eq:Il-dec}
\end{equation}
\begin{equation}
I_{l,\pm}(\tr,\tau):=\frac{1}{2i}
\int\limits_0^{+\infty} {\rm d}\eta\ [\sinh (2\eta)]^2\ 
e^{-4l\eta}\,e^{-il \tau\sinh (2\eta)\pm i
\tr\sinh\eta}~.\label{eq:Ilpm-def}
\end{equation}

The integrals $I_{l,\pm}$ must be treated differently. 
Specifically, the major contribution to integration in $I_{l,-}$ always arises from the neighborhood
of the endpoint, $\eta=0$: \looseness=-1
\begin{equation}
I_{l,-}(\tr,\tau)\sim  \frac{4}{(2l\, \tau+ \tr-i\,4l)^3}\sim \frac{4}{(2l\,\tau+\tr)^3}\qquad \mbox{all}\quad \tr\ge 0~.\label{eq:Il-}
\end{equation}
It follows that~\cite{batemanI}
\begin{equation}
\sum_{l=1}^{+\infty}I_{l,-}(\tr,\tau)\sim -\frac{1}{4\tau^3}\,\psi^{\prime\prime}\biggl(1+\frac{\tr}{2\tau}\biggr)~,
\label{eq:Il-sum}
\end{equation}
where $\psi^{\prime\prime}(z):={\rm d}^2\psi(z)/{\rm d}z^2$.

By contrast, computing $I_{l,+}$ requires the evaluation of a stationary-phase contribution 
when $\tr$ is comparable to $\tau$. Most of the remaining subsection is devoted to the analytical
computation of $I_{l,+}$.\looseness=-1
\vskip10pt

\noindent{\bf Remarks on the integral $I_{l,+}$.}
To evaluate $I_{l,+}$, it is advisable to introduce the associated phase 
\begin{equation}
\Theta(\eta):=(\tr-2l\tau\cosh \eta)\sinh\eta~.\label{eq:Theta-def}
\end{equation}
The stationary-phase points are the non-negative roots of the equation ${\rm d}\Theta/{\rm d}\eta =0$:
\begin{equation}
\tr\cosh\eta-2l \tau\cosh(2\eta)=0~,\label{eq:Theta-pr}
\end{equation}
which is solved by $\eta=\eta_l\ge 0$ where
\begin{equation}
\cosh\eta_{l}=\frac{\beta_l +\sqrt{\beta_l^2+8}}{4}\ge 0\qquad\mbox{if}\quad
\beta_l:=\frac{\tr}{2l \tau}\ge 1~.\label{eq:ss-def}
\end{equation}
Note that for $\eta_l >0$, 
\begin{equation}
\Theta(\eta_l)>0,\ \Theta^{\prime\prime}(\eta_l):=
{\rm d}^2\Theta/{\rm d}\eta^2|_{\eta_l}=(\tr-8l\, \tau\cosh\eta_l)\sinh\eta_l<0~.\label{eq:Theta-ineq}
\end{equation}
In particular, if $\eta_l\ge O(1)$ then $\Theta(\eta_l)\ge O(l\,\tau)$ and 
$|\Theta^{\prime\prime}(\eta_l)|\ge O(l\, \tau)$. In this case, the stationary-phase
calculation is carried out as usual~\cite{cheng}; see~(\ref{eq:Il+-stph}) below.

However, caution should be exercised in the asymptotic evaluation of $I_{l,+}(\tr,\tau)$ when the point
$\eta_l=\eta_l(\tr,\tau)$ lies too close to $\eta=0$. This coalescence occurs when
$\eta_l$ becomes of the order of $|\Theta''(\eta_l)|^{-1/2}$ while $0\le\beta_l-1\ll 1$. 
By expanding $\eta_l$ near $\beta_l =1$, we find that a more precise condition for this coalescence is
\begin{equation}
|2l\, \tau-\tr|=O\big((l\tau)^{1/3}\big)~.\label{eq:coals}
\end{equation}Accordingly, we distinguish three main regions for ($\tr,\tau$) as shown below. 
In the end of this section,
we exploit the overlapping asymptotic expansions for $I_l$ in these regions in order to construct 
a connection formula.
\vskip10pt

\noindent{\bf Region I: $2l\, \tau-\tr\gg (l \tau)^{1/3}$.} 
In this case, the stationary-phase point $\eta_l$ is not real.
The major contribution to $I_{l,+}$ stems from $\eta=0$:
\begin{eqnarray}
I_{l,+}(\tr,\tau) &\sim&  \frac{1}{2i}\int\limits_0^{+\infty} {\rm d}\eta\
(2\eta)^2\,e^{-i(2l \tau-\tr)\eta}\,e^{-4l\eta}\nonumber\\
{} &=& \frac{4}{(2l \tau-\tr-i\,4l)^3}\sim \frac{4}{(2l \tau-
\tr)^3}~.\label{eq:Ilp-rs}
\end{eqnarray}
This result, combined with~(\ref{eq:Il-dec}) and~(\ref{eq:Il-}), 
furnishes (\ref{eq:Il-app}) when $\tr\ll 2l\tau$. 

We note in passing that if $\tr\le  O(2l\tau)$ yet $2l\tau -\tr\gg (l\tau)^{1/3}$ 
for all $l$ then we can sum up~(\ref{eq:Ilp-rs}). Consequently,
\begin{equation}
\sum_{l=1}^{+\infty}I_{l,+}(\tr,\tau)\sim -\frac{1}{4\tau^3}\,\psi^{\prime\prime}
\biggl(1-\frac{\tr}{2\tau}\biggr)~.
\label{eq:Il+-sum}
\end{equation}
Formulas~(\ref{eq:Il-sum}) and~(\ref{eq:Il+-sum}) combined yield
\begin{equation}
\Lambda(r,t)\sim \frac{1}{8\pi^2\,\tr}\,\frac{1}{\tau^3}\,
\biggl[\psi^{\prime\prime}\biggl(1+\frac{\tr}{2\tau}\biggr)-\psi^{\prime\prime}
\biggl(1-\frac{\tr}{2\tau}\biggr)\biggr]~.\label{eq:Lambda-r-sm}
\end{equation}
This formula reduces to~(\ref{eq:Lambda-r-fix}) via the Taylor expansions of 
$\psi^{\prime\prime}\big(1\pm \tr/(2\tau)\big)$ about $1$.
\vskip10pt

\noindent{\bf Region II: $\tr-2l\, \tau\gg (l\, \tau)^{1/3}$.} 
The standard method of stationary phase is now applicable.
The integral $I_{l,+}$ is approximated by~\cite{cheng}
\begin{equation}
I_{l,+}(\tr,\tau)\sim -\sqrt{\frac{\pi}{2 |\Theta^{\prime\prime}(\eta_l)|}}\ e^{i
\Theta(\eta_l)+i\pi/4}
\ [\sinh (2\eta_l)]^2\ e^{-4l\eta_l}~,\label{eq:Il+-stph}
\end{equation}
where $\Theta(\eta_l)$ and $\Theta^{\prime\prime}(\eta_l)$ are given 
by~(\ref{eq:Theta-def}) and~(\ref{eq:Theta-ineq}).

It is of interest to take the limit of (\ref{eq:Il+-stph})
as $\beta_l\to 1$, i.e. $\eta_l\to 0$. With
\begin{eqnarray}
\Theta(\eta_l) &\sim&   2l \tau\,\beta_l\, (\eta_l
+{\textstyle\frac{1}{6}}\eta_l^3)-2l\,
\tau(\eta_l+{\textstyle\frac{2}{3}}\eta_l^3)\nonumber\\ 
{} &\sim & \frac{2^{5/2}}{3^{3/2}}\,(\beta_l-1)^{3/2}\,l \tau~,\label{eq:Theta-exp1}
\end{eqnarray}
\begin{equation}
\Theta''(\eta_l)\sim -2^{3/2}\,3^{1/2}\,(\beta_l-1)^{1/2}\,l \tau\quad
\hbox{as $\beta_l\to 1^+$}~,\label{eq:Thetap-exp1}
\end{equation}
the stationary-phase formula (\ref{eq:Il+-stph}) reduces to
\begin{equation}
I_{l,+}(\tr,\tau)\sim -\frac{2^{7/4}}{3^{5/4}}
\sqrt{\frac{\pi}{l \tau}}\, (\beta_l-1)^{3/4}
\exp\biggl[i\frac{2^{5/2}}{3^{3/2}}\,(\beta_l-1)^{3/2}\,l\tau+i\frac{\pi}{4}\biggr].\label{eq:Il+_exp1}
\end{equation}
\vskip10pt

\noindent{\bf Region III: $|2l\, \tau-\tr|= O\big((l\,\tau)^{1/3}\big)$.} In this more demanding case, the stationary-phase
point $\eta_l$ is too close to the endpoint of integration and $\Theta(\eta)$ must be expanded at $\eta=0$.
Approximation~(\ref{eq:Theta-exp1}) indicates that terms $O(\eta^3)$ must be retained in $\Theta(\eta)$.

First, we consider the region $2l\tau > \tr$ with $2l\tau-\tr=O\big((l\tau)^{1/3}\big)$. With the definition
\begin{equation}
\gamma_l:=\frac{2}{3^{3/2}}\,\frac{(2l \tau-\tr)^{3/2}}{(l
\tau)^{1/2}}\label{eq:tild-beta}
\end{equation}
and the change of variable $\eta=2(l\tau)^{-1/3}(\gamma_l/2)^{1/3}\,\sinh(v/3)$ in~(\ref{eq:Ilpm-def}), 
$I_{l,+}$ becomes
\begin{eqnarray}
I_{l,+}(\tr,\tau)&\sim&  \frac{1}{2i}\int\limits_0^{+\infty}{\rm d}\eta\ (2\eta)^2\,e^{-4l\eta}\,
e^{i(\tr-2l\tau)\eta-il\,\tau\eta^3}\nonumber\\
{} &\sim& \frac{2}{3il \tau}\,
\gamma_l\int\limits_0^{+\infty}{\rm d}v\, [\cosh v-\cosh(v/3)]\,
e^{-i\,\gamma_l\,\sinh v}\nonumber\\
{} &=&   -\frac{2}{3l \tau}+\frac{4i}{3}\,\left(\frac{2l \tau-\tr}{3l
\tau}\right)^{3/2}\,S_{0,\frac{1}{3}}(i\gamma_l)~,\quad \gamma_l=O(1)~,
\label{eq:Il+_Lom1}
\end{eqnarray}
where $S_{\mu,\nu}$ is the Lommel function~\cite{bateman}.
In the above, the $e^{-4l\,\eta}$ term has been
neglected in the integrand. This simplification is adequate if, for instance,
$\tr=O(\tau)$ and only values $l=O(1)$ are of interest. 
Note that if $2l \tau\to \tr$ for some $l$ then $\gamma_l\to 0$ and the second term
in (\ref{eq:Il+_Lom1}) vanishes, as it should.\looseness=-1

Formula~(\ref{eq:Il+_Lom1}) connects
smoothly to approximation (\ref{eq:Ilp-rs}) when $\gamma_l\gg 1$.
Indeed, by use of the asymptotic formula~\cite{bateman}
\begin{equation}
S_{0,\frac{1}{3}}(z)\sim \frac{1}{z}-\frac{8}{9z^3}\quad \hbox{as
$|z|\to +\infty$}~,
\quad |{\rm Arg}\ z|<\pi~,\label{eq:Lom-asymp}
\end{equation}
the first term in (\ref{eq:Il+_Lom1}) is exactly canceled.
Thus,~(\ref{eq:Il+_Lom1}) reduces to (\ref{eq:Ilp-rs}).

To account for large values of $l$, $l=O(\tau^{1/2})$ by which $2l\, \tau-\tr=O(l)$, the $e^{-4l\eta}$
factor has to be retained in the integral~(\ref{eq:Ilpm-def}) for $I_{l,+}$. An inspection of
(\ref{eq:Il+_Lom1}) shows that $\gamma_l$ has to be replaced by
\begin{equation}
\tilde\gamma_l:= \frac{2}{3^{3/2}}\,\frac{(2l \tau-\tr-i\,4l)^{3/2}}{(l\tau)^{1/2}}~.\label{eq:brbeta-def}
\end{equation}
The ensuing approximation for $I_{l,+}$ in place of~(\ref{eq:Il+_Lom1}) reads
\begin{equation}
I_{l,+}(\tr,\tau)\sim -\frac{2}{3l\tau}+\frac{4i}{3}\,\left(\frac{2l
\tau-\tr-i\,4l}{3l\tau}\right)^{3/2}\,S_{0,\frac{1}{3}}(i\tilde\gamma_l)~.\label{eq:Il+_Lom2}
\end{equation}

Next, we restrict attention to the region $\tr > 2l\tau$ with $\tr-2l\tau= O\big((l\,\tau)^{1/3}\big)$. This case
is not essentially different from the previous one. Formula
(\ref{eq:Il+_Lom1}) can be continued analytically to complex values of $\gamma_l$,
as $\gamma_l$ varies continuously from $\gamma_l=|\gamma_l|$ for $\tr< 2l\tau$
to $|\gamma_l|e^{-i3\pi/2}$ or $|\gamma_l|e^{i3\pi/2}$ when $\tr > 2l\tau$.
Both continuations should yield the same result for $I_{l,+}$ since
this integral is a single-valued function of $\gamma_l$. Indeed, from~\ref{app:C} 
we have  
\begin{eqnarray}
\lefteqn{(\gamma_l\,e^{-i3\pi/2})\,
[(i\gamma_l\,e^{-i3\pi/2})^{-1}-S_{0,\frac{1}{3}}(i\,\gamma_l\,
e^{-i3\pi/2})]}\nonumber\\
{}\mbox{}\quad  &=&   -i-i\,\gamma_l\,S_{0,\frac{1}{3}}(\gamma_l\,e^{-i\pi})
=-i+i\,\gamma_l\, S_{0,\frac{1}{3}}(\gamma_l\,e^{i2\pi})\nonumber\\ 
{}\mbox{}\quad  &=&   (\gamma_l\,e^{i3\pi/2})\,
[(i\gamma_l\,e^{i3\pi/2})^{-1}-S_{0,\frac{1}{3}}(i\,\gamma_l\,
e^{i3\pi/2})]~.\label{eq:analt-cont}
\end{eqnarray}

Equation (\ref{eq:Il+_Lom1}) combined with~(\ref{eq:Hankel-I}) of~\ref{app:C} yields
\begin{eqnarray}
I_{l,+}(\tr,\tau)  &\sim&    -\frac{2}{3l
\tau}\,|\gamma_l|\,[|\gamma_l|^{-1}+S_{0,\frac{1}{3}}(|\gamma_l|\,e^{-i\pi})]\nonumber\\
{}  &=&    -\frac{2}{3l
\tau}+\frac{4}{3}\,\left(\frac{\tr-2l\tau}{3l\tau}\right)^{3/2}\nonumber\\
&&  \mbox{} \times\left[S_{0,\frac{1}{3}}(|\gamma_l|)
+\frac{\pi}{2}\,\sqrt{3}\,e^{-i\pi/3}\,H_{1/3}^{(1)}
(|\gamma_l|)\right]~,\quad \tr > 2l\tau~,\label{eq:Il+_Lom3}
\end{eqnarray}
where, by~(\ref{eq:tild-beta}), $|\gamma_l|=2\cdot 3^{-3/2}(\tr-2l\tau)^{3/2}(l\tau)^{-1/2}$.

For $|\gamma_l|\gg 1$ and $\tr\to 2l\tau$ (i.e., $\beta_l\to 1$), we show that~(\ref{eq:Il+_Lom3}) connects smoothly to the
stationary-phase contribution of~(\ref{eq:Il+_exp1}). The leading term in the brackets of~(\ref{eq:Il+_Lom3})
arises from $H_{1/3}^{(1)}$ according to the expansion~\cite{bateman}\looseness=-1
\begin{equation}
H_{1/3}^{(1)}(x)\sim \sqrt{\frac{2}{\pi
x}}\ e^{i(x-5\pi/12)}\,[1+O(x^{-1})]\qquad x\to +\infty~.\label{eq:H1-asymp}
\end{equation}
The next asymptotic term in~(\ref{eq:Il+_Lom3}) comes from the $S_{0,\frac{1}{3}}$ and
exactly cancels the $-2/(3l\tau)$. Thus, (\ref{eq:Il+_Lom3}) reduces to\looseness=-1
\begin{equation}
I_{l,+}(\tr,\tau)\sim -\frac{2}{3^{5/4}}\,\sqrt{\frac{\pi}{l \tau}}\,
\left(\frac{\tr-2l\tau}{l\tau}\right)^{3/4}\,
e^{i\,|\gamma_l|+i\pi/4}~.\label{eq:Il+_exp4}
\end{equation}
This formula is in agreement with~(\ref{eq:Il+_exp1})
under the substitutions $\tr=\beta_l\,(2l \tau)$ and
\begin{equation}
|\gamma_l|=2^{5/2} 3^{-3/2}\,(\beta_l-1)^{3/2}\,l\tau\sim
\Theta(\eta_l)\qquad \eta_l\to 0~.\label{eq:gamml-com}
\end{equation}
The inclusion of the $e^{-4l\,\eta}$ factor in~(\ref{eq:Ilpm-def}) for $I_{l,+}$
necessarily modifies (\ref{eq:Il+_Lom3}). By defining
\begin{eqnarray}
\breve\gamma_l &:=&   \frac{2}{3^{3/2}}\,\frac{(\tr-2l \tau+i\,4l)^{3/2}}{(l \tau)^{1/2}}\label{eq:brevgamm}\\
{} &\sim&  |\gamma_l|\,[1+i\,(8/3)l\eta_l\,|\gamma_l|^{-1}]^{3/2}~,\quad
0\le\eta_l\ll 1~,\label{eq:brevgam-app}
\end{eqnarray}
we replace~(\ref{eq:Il+_Lom3}) by the formula
\begin{eqnarray}
I_{l,+}&\sim& -\frac{2}{3l\,\tau}+\frac{4}{3}\,\left(\frac{\tr-2l\tau+i\,4l}{3l
\tau}\right)^{3/2}\nonumber\\
\mbox{}&&\times \left[S_{0,\frac{1}{3}}(\breve\gamma_l)
+\frac{\pi}{2}\,\sqrt{3}\,e^{-i\pi/3}\,H_{1/3}^{(1)}(\breve\gamma_l)\right]~.\label{eq:Il+_exp5}
\end{eqnarray}
\vskip10pt

\noindent{\bf Connection formula for $I_{l,+}$.}  
Next, we derive an asymptotic formula that connects the apparently
disparate yet overlapping formulas~(\ref{eq:Ilp-rs}), (\ref{eq:Il+-stph}) and~(\ref{eq:Il+_Lom1}) for $I_{l,+}$.
It is instructive, although not necessary, to distinguish the cases $\tr< 2l\tau$ and $\tr\ge 2l\tau$.

Assume that $\tr< 2l\tau$. By virtue of~(\ref{eq:Il+_Lom2}),
$I_{l,+}(\tr,\tau)$ reads
\begin{eqnarray}
I_{l,+}(\tr,\tau)  &\sim &  
-\frac{2}{3l\tau}+\frac{4i}{3}\left(\frac{2l\tau-\tr-i4l}{3l\tau}\right)^{3/2}\nonumber\\
\mbox{}&& \qquad\times S_{0,\frac{1}{3}}\left(\frac{2i}{3}\,\frac{(2l\tau-\tr-i4l)^{3/2}}{(3l\tau)^{1/2}}\right)~.\label{eq:Il-1}
\end{eqnarray}
This formula can be continued analytically to the region $\tr> 2l\tau$ as was shown above, but may
break down when $\tr- 2l\tau\gg O\big((l\tau)^{1/3}\big)$ (in region~II).
Our task is to connect this formula with the stationary-phase result~(\ref{eq:Il+-stph})
by using a single expression.

Let us now consider $\tr> 2l\tau$. 
By inspection of~(\ref{eq:Il+_Lom3}), (\ref{eq:Il+_exp4}), (\ref{eq:gamml-com}) 
and~(\ref{eq:brevgam-app}), we propose
to start with the composite expression
\begin{equation}
I_{l,+}\sim -{\mathcal L}\,\frac{2}{3l
\tau}-{\mathcal C}\,S_{0,\frac{1}{3}}\left(\Theta(\eta_l)[1+i(8/3)l\eta_l
\Theta(\eta_l)^{-1}]^{3/2}e^{-i\pi}\right),\label{eq:Il+_comp} 
\end{equation}
where the constants ${\mathcal L}$ and ${\mathcal C}$ are to be determined. This
formula must connect smoothly to 
the stationary-phase result~(\ref{eq:Il+-stph}) when $\eta_l\ge O(1)$ and, thus, $\Theta(\eta_l)\gg 1$. 
Recall that the phase function $\Theta(\eta)$ is defined by~(\ref{eq:Theta-def}).

In order to find ${\mathcal L}$ and ${\mathcal C}$ we replace $S_{0,\frac{1}{3}}$ in~(\ref{eq:Il+_comp}) 
by its large-argument expansion, and then match the asymptotic result with~(\ref{eq:Il+-stph}).
By~\ref{app:C}, we use the relation~\cite{bateman} 
\begin{eqnarray}
S_{0,\frac{1}{3}}(z\,e^{-i\pi})  &=&   -S_{0,\frac{1}{3}}(z)-
\frac{\pi}{2}\,\sqrt{3}\,e^{-i\pi/3}\,H_{1/3}^{(1)}(z)\nonumber\\[3pt]
{}  &\sim&    \sqrt{\frac{3\pi}{2 z}}\, e^{i z+i\pi/4}-\frac{1}{z},\quad
|z|\to\infty,\ |{\rm Arg}\ z|<\pi~.\label{eq:SH-asymp}
\end{eqnarray}
For $\eta_l\ge O(1)$, $I_{l,+}$ is thus approximated by
\begin{eqnarray}
\lefteqn{I_{l,+}(\tr,\tau)  \sim   
-{\mathcal L}\,\frac{2}{3l \tau}+\frac{{\mathcal C}}{\Theta(\eta_l)+i4l\,\eta_l}-{\mathcal C}\,
\sqrt{\frac{3\pi}{2[\Theta(\eta_l)+i4l\,\eta_l]}}}\nonumber\\
\mbox{}&&\qquad \times e^{i\Theta(\eta_l)+i\pi/4}\,e^{-4l\,\eta_l}\nonumber\\
{}  &\sim&    -\frac{2}{3l \tau}\,\left[{\mathcal L}-{\mathcal C}\,\frac{3}{2}\, \frac{l
\tau}{\Theta(\eta_l)}\right] -{\mathcal C}\,\sqrt{\frac{3\pi}{2\Theta(\eta_l)}}\,
e^{i\Theta(\eta_l)+i\pi/4}\,e^{-4l\eta_l}~.\label{eq:Il+_comp2}
\end{eqnarray}
The term in the brackets of~(\ref{eq:Il+_comp2}) should be zero while the second term
must be identified with (\ref{eq:Il+-stph}). Accordingly, we have the relations
\begin{equation}
{\mathcal L}={\mathcal C}\,\frac{3}{2}\, \frac{l\tau}{\Theta(\eta_l)}~,\label{eq:LN-1}
\end{equation}
\begin{equation}
{\mathcal C}=\sqrt{\frac{\Theta(\eta_l)}{3\,|\Theta''(\eta_l)|}}\,[\sinh
(2\eta_l)]^2~.\label{eq:LN-2}
\end{equation}
By use of the identity 
\begin{equation}
\frac{\Theta(\eta_l)}{|\Theta''(\eta_l)|}=
\frac{\tr\sinh\eta_l-l\tau\sinh (2\eta_l)}{4l\tau\sinh (2\eta_l)-
\tr\sinh\eta_l}=\frac{(\sinh\eta_l)^2}{1+2(\cosh\eta_l)^2}~,\label{eq:Theta-id}
\end{equation}
we find
\begin{equation}
{\mathcal C}=\frac{1}{\sqrt{3}}\,\frac{\sinh\eta_l}{\sqrt{1+2(\cosh\eta_l)^2}}\,
[\sinh(2\eta_l)]^2~,\label{eq:N-fin}
\end{equation}
\begin{equation}
{\mathcal L}=\sqrt{3}\,\frac{(\cosh\eta_l)^3}{\sqrt{1+2(\cosh\eta_l)^2}}~.\label{eq:L-fin}
\end{equation}

To further validate~(\ref{eq:Il+_comp}) given~(\ref{eq:N-fin})
and~(\ref{eq:L-fin}), we consider $\tr-2l\tau=O\big((l\tau)^{1/3}\big)$ (in region~III) so that
$0\le \eta_l\ll 1$. In this limit, (\ref{eq:N-fin}) and~(\ref{eq:L-fin}) entail 
${\mathcal C}\sim (4/3)\eta_l^3\sim (2/3)|\gamma_l|$ and ${\mathcal L}\sim 1$, while the argument
of $S_{0,\frac{1}{3}}$ in~(\ref{eq:Il+_comp}) becomes approximately $|\gamma_l|e^{-i\pi}$.
Recall the definition~(\ref{eq:gamml-com}) of $\gamma_l$.
Thus,~(\ref{eq:Il+_comp}) reduces to~(\ref{eq:Il+_Lom3}).

We point out a refinement of our procedure.
A comparison of~(\ref{eq:Il+_comp}) with formula (\ref{eq:Il+_exp5}) reveals that the
${\mathcal C}$ of~(\ref{eq:N-fin}) does not reproduce the expected prefactor of Lommel's
function when $\tr-2l\, \tau= O(l)$, i.e. $\eta_l= O(\tau^{-1/2})$. A  remedy to this discrepancy
is to replace $\eta_l$ in the ${\mathcal C}$ of~(\ref{eq:N-fin}) by
\begin{equation}
\bar\eta_l:=\sqrt{\eta_l^2+\frac{4i}{3 \tau}}~.\label{eq:bareta-def}
\end{equation}
This $\bar\eta_l$ stems from including the exponent 
$-4l\eta$ in definition~(\ref{eq:Theta-def}) for 
$\Theta(\eta)$. The equation for the resulting, modified stationary-phase point $\breve\eta_l$ reads
\begin{equation}
2l \tau\cosh (2\breve\eta_l)-\tr\cosh\breve\eta_l-i4l=0~,\label{eq:stph-mod}
\end{equation}
which has solution
\begin{equation}
\cosh\breve\eta_l=\frac{\beta_l+\sqrt{\beta_l^2+8+i\,16/\tau}}{4}~.\label{eq:etal-mod}
\end{equation}
Expanding this equation about $\beta_l=1$ yields
\begin{eqnarray}
1+\frac{\breve\eta_l^2}{2}  &\sim&  
1+\frac{\beta_l-1+i2/\tau}{3}=1+\frac{1}{3}\,
\frac{\tr+i4l-2l\, \tau}{2l\tau}~,\label{eq:etal-refn}
\end{eqnarray}
by which $\breve\eta_l\sim \bar\eta_l$ in view of~(\ref{eq:bareta-def}).

A connection formula for $I_{l,+}$ follows from~(\ref{eq:Il+_comp}) with~(\ref{eq:N-fin})--(\ref{eq:bareta-def}):
\begin{eqnarray}
I_{l,+}(\tr,\tau)  &\sim&   
-\frac{2}{\sqrt{3}}\,\frac{(\cosh\eta_l)^3}{\sqrt{1+2(\cosh\eta_l)^2}}\,\frac{1}{l\,\tau}
-\frac{1}{\sqrt{3}}\,\frac{(\sinh\bar\eta_l)\,[\sinh(2\bar\eta_l)]^2}{\sqrt{1
+2(\cosh\eta_l)^2}}\nonumber\\ 
&&  \mbox{} \times
S_{0,\frac{1}{3}}\left(\Theta(\eta_l)[1+i(8/3)l\eta_l\,
\Theta(\eta_l)^{-1}]^{3/2}\,e^{-i\pi}\right)~.\label{eq:Il+_conn}
\end{eqnarray}
\vskip10pt

\noindent{\bf Asymptotic formula for $\Lambda(r,t)$.} A large-$t$ asymptotic formula for 
$\Lambda(r,t)=(16\pi a\rho_0)^{-3/2}[\mathcal K({\bf r},t)-\mathcal K({\bf r},+\infty)]$ is obtained 
by~(\ref{eq:Lambda-I}), (\ref{eq:Il-dec}), (\ref{eq:Il-sum}) and~(\ref{eq:Il+_conn}):
\begin{eqnarray}
\lefteqn{\Lambda(r,t)\sim -\frac{1}{2\pi^2 \tr}\Biggl\{\sum_{l=1}^{[\tr/(2 \tau)]}
\Biggl[\frac{(\cosh\eta_l)^3}{\sqrt{1+2(\cosh\eta_l)^2}}\,
\frac{2}{\sqrt{3}\,l\tau}}\nonumber\\  
&&  \mbox{}
+\frac{1}{\sqrt{3}}\,\frac{(\sinh\bar\eta_l)\,[\sinh(2\bar\eta_l)]^2}{\sqrt{1+2(\cosh\eta_l)^2}}\nonumber\\ 
&&  \mbox{} \times S_{0,\frac{1}{3}}\left(\Theta(\eta_l)[1+i(8/3)l\eta_l\,
\Theta(\eta_l)^{-1}]^{3/2}\,e^{-i\pi}\right)\Biggr]\nonumber\\
&&  \mbox{} +\sum_{l=[\tr/(2\tau)]+1}^{+\infty}\Biggl[\frac{2}{3l
\tau}-\frac{4i}{3}\,\left(\frac{2l \tau-\tr-i\,4l}{3l \tau}\right)^{3/2}\nonumber\\
&&  \mbox{} \times
S_{0,\frac{1}{3}}\left(\frac{2i}{3}\,\frac{(2l\tau-\tr-i\,4l)^{3/2}}{(3l\tau)^{1/2}}
\right) \Biggr]-\frac{1}{4\tau^3}\,\psi''(1+\tr/(2\tau))\Biggr\}~,\label{eq:Lambda-asymp}
\end{eqnarray}
where $[x]$ denotes the largest integer that is less than or equal to $x$. The infinite series for 
$\Lambda$ is absolutely convergent and can be simplified depending on the value of $\tr/(2\tau)$.
For instance, if $\tr < 2\tau$ then (\ref{eq:Lambda-asymp}) reduces approximately
to the expression
\begin{eqnarray}
\Lambda(r,t)&\sim& -\frac{1}{2\pi^2\,\tr}\Biggl\{\frac{2}{3\tau}-\frac{4i}{3}
\Biggl(\frac{2\tau-\tr-4i}{3\tau}\Biggr)^{3/2}\,S_{0,\frac{1}{3}}\Biggl(\frac{2i}{3}\frac{(2\tau-\tr-4i)^{3/2}}{(3\tau)^{1/2}}\Biggr)
\nonumber\\
\mbox{} && -\frac{1}{4\tau^3}\Biggl[\psi^{\prime\prime}\Biggl(1+\frac{\tr}{2\tau}\Biggr)
-\psi^{\prime\prime}\Biggl(1-\frac{\tr}{2\tau}\Biggr)-2\Biggl(1-\frac{\tr}{2\tau}\Biggr)^{-3}\Biggr]\Biggr\}~.
\label{eq:Lambda-asymp-simpl}
\end{eqnarray}
The last formula shows that the magnitude of $\Lambda$ varies from $O(\tau^{-4})$ when
$\tr=O(1)$ to $O\big((\tr\tau)^{-1}\big)$ as $r$ approaches $2\tau$.

More generally, if $2(n-1)\tau < \tr < 2n\tau$ for some positive integer $n=O(1)$ then~(\ref{eq:Lambda-asymp})
is approximated by the following finite sums:\looseness=-1
\begin{eqnarray}
\lefteqn{\Lambda(r,t) \sim -\frac{1}{2\pi^2 \tr}\Biggl\{\sum_{l=1}^{n-1}
\Biggl[\frac{(\cosh\eta_l)^3}{\sqrt{1+2(\cosh\eta_l)^2}}\,
\frac{2}{\sqrt{3}\,l\tau}}\nonumber\\  
&&  \mbox{}
+\frac{1}{\sqrt{3}}\,\frac{(\sinh\bar\eta_l)\,[\sinh(2\bar\eta_l)]^2}{\sqrt{1+2(\cosh\eta_l)^2}}\nonumber\\ 
&&  \mbox{} \times S_{0,\frac{1}{3}}\left(\Theta(\eta_l)[1+i(8/3)l\eta_l\,
\Theta(\eta_l)^{-1}]^{3/2}\,e^{-i\pi}\right)\Biggr]\nonumber\\
&& \mbox{} +\frac{2}{3n\tau}-\frac{4i}{3}\Biggl(\frac{2n\tau-\tr-4i\,n}{3n\tau}\Biggr)^{3/2}
S_{0,\frac{1}{3}}\Biggl(\frac{2i}{3}\frac{(2n\tau-\tr-4i\,n)^{3/2}}{(3n\tau)^{1/2}}\Biggr)\nonumber\\
&&\mbox{} -\frac{1}{4\tau^3}\Biggl[\psi^{\prime\prime}\Biggl(1+\frac{\tr}{2\tau}\Biggr)-
\psi^{\prime\prime}\Biggl(1-\frac{\tr}{2\tau}\Biggr)-2\sum_{l=1}^n\Biggl(l-\frac{\tr}{2\tau}\Biggr)^{-3}
\Biggr]\Biggr\}~.\label{eq:Lambda-asymp-n}
\end{eqnarray}
As $\tr/(2\tau)$ approaches $n$ the magnitude of $\Lambda$ locally increases to become
$O\big((n\,\tr\tau)^{-1}\big)$.

Recall that the dependence on $r$, $t$ and $a\rho_0$ is restored via definitions~(\ref{eq:tau-r})
by which $\tr=(16\pi a\rho_0)^{1/2} r$ and $\tau=(16\pi a\rho_0)t$.
Note that the quantity $(16\pi a\rho_0)^{1/2}$ is the sound velocity in units
where $\hbar=1=2m$~\cite{leehuangyang}.
Equation~(\ref{eq:Lambda-asymp}) is further discussed in section~\ref{sec:conclusion}.

The asymptotic analysis of this section can be extended to nonzero spherically
symmetric initial data, $f_0(r)\neq 0$, by which $p_0(k)\neq \widehat g_0(k)$.
The major contribution to the requisite integral~(\ref{eq:hatK0-exp})
still comes from the vicinity of $k=0$. For example, for $l=O(1)$ we can apply the approximation
$[p_0(k)\,\widehat g_0(k)]^l\sim [p_0(0)\,\widehat g_0(0)]^l$ if $p_0(0)\,\widehat g_0(0)\neq 0$.

\section{Slowly varying trapping potential}
\label{sec:slow}

In this section we use a potential
$V_e({\bf x})$ in~(\ref{eq:Phi-NLSE}) and~(\ref{eq:K0-IDE-app}) that is slowly varying and 
increases with $|{\bf x}|$. 
We invoke the approximate time-independent wavefunction $\Phi({\bf x})$ for the
macroscopic state derived in~\cite{wu98}. Accordingly, 
we obtain the time-dependent pair-excitation function $K_0$ by solving~(\ref{eq:K0-IDE-app}).
We show that, under certain conditions, the analysis for $K_0$ is not essentially different from the
procedure of section~\ref{sec:trans-inv}.\looseness=-1

\subsection{Macroscopic wavefunction}
\label{subsec:phi-app}

Here, we revisit briefly the solution of the nonlinear Schr\"odinger equation~(\ref{eq:Phi-NLSE})
that was studied in~\cite{wu98}. 

By the replacement $\Phi({\bf x},t)=e^{-iEt}\Phi({\bf x})$, the equation for $\Phi({\bf x})$ is
\begin{equation}
[-\Delta_x+V_e({\bf x})+8\pi a\rho_0 \Phi({\bf x})^2-4\pi a\rho_0\zeta]\Phi({\bf x})=E\,\Phi({\bf x})~,
\label{eq:Phi-tind}
\end{equation}
where $E$ is the energy per particle of the macroscopic state and $\Phi({\bf x})$ is chosen to be real.
We consider a sufficiently slowly varying $V_e(x)$, formally described by 
$V_e(x)=:\tilde V_e(\epsilon x)$ where $0<\epsilon \ll 1$ and $\tilde V_e=O(1)$. Accordingly,
$\Delta_x\Phi$ is neglected in~(\ref{eq:Phi-tind}). Thus, $\Phi\sim \Phi_0$ satisfies the equation~\cite{wu98}
\begin{equation}
[V_e({\bf x})+8\pi a\rho_0\Phi_0({\bf x})^2-4\pi a\rho_0\zeta-E]\Phi_0({\bf x})=0~.\label{eq:Phi-noL}
\end{equation}
The solution of this algebraic equation reads~\cite{wu98}
\begin{equation}
\Phi_0({\bf x})=\left\{\begin{array}{lr} (8\pi a\rho_0)^{-1/2}[4\pi a\rho_0\zeta+E-V_e({\bf x})]^{1/2}~, & {\bf x}\in \mathcal R~,\\
                                                                                                      0~, & {\bf x}\notin \mathcal R~,
                        \end{array}\right.~\label{eq:Phi-app}
\end{equation}
where the region $\mathcal R$ is defined by~\cite{wu98}
\begin{equation}
\mathcal R=\{ {\bf x}:\ 4\pi a\rho_0\zeta+ E > V_e({\bf x})\}~.
\label{eq:regR-def}
\end{equation}
The function $\Phi_0$ must vanish as ${\bf x}$ approaches the boundary, $\partial\mathcal R$,
 of $\mathcal R$ from the interior of $\mathcal R$.
Notice that $\Phi_0$ by~(\ref{eq:Phi-app}) is also slowly varying.
The multiplication of~(\ref{eq:Phi-app}) by $\Phi_0({\bf x})$ and subsequent integration in ${\bf x}$ yield~\cite{wu98}
\begin{equation}
E=4\pi a\rho_0\zeta+\zeta_e~.\label{eq:E-zeta}
\end{equation}

The right-hand side of~(\ref{eq:Phi-app}) is not differentiable in ${\bf x}$ 
at $\partial\mathcal R$. 
A remedy to this problem was provided in~\cite{wu98} by retainment
of the Laplacian in~(\ref{eq:Phi-tind}).

\subsection{Pair-excitation function}
\label{subsec:K-app}

In this subsection we determine the time-dependent pair-excitation function in the center-of mass coordinates 
under approximation~(\ref{eq:Phi-app}). In view of integrodifferential equation~(\ref{eq:K0-com}), let
\begin{equation}
\mathcal K({\bf R},{\bf r},t)=e^{-i2Et}\,\mathcal K_0({\bf R},{\bf r},t)~.\label{eq:K0-E}
\end{equation}

In light of the translationally invariant case (section~\ref{sec:trans-inv}), 
we treat ${\bf R}$ and ${\bf r}$ as slow and fast variables,
respectively, in~(\ref{eq:K0-com}). This consideration amounts to neglecting the Laplacian $\Delta_R$ 
and eliminating the variable ${\bf r}$ in the arguments of $V_e$ and $\Phi$ in~(\ref{eq:K0-com}). In
addition, we have~\cite{wu98}
\begin{equation}
{\mathcal K}_0({\bf R}+\textstyle{\frac{1}{2}}{\bf r}-\textstyle{\frac{1}{2}}{\bf w},{\bf w})
{\mathcal K}_0({\bf R}-\textstyle{\frac{1}{2}}{\bf w},{\bf w}-{\bf r})
\sim \mathcal K_0({\bf R},{\bf w})\mathcal K_0({\bf R},{\bf w}-{\bf r}).\label{eq:K0-intg}
\end{equation}
Thus,~(\ref{eq:K0-com}) reduces to~\cite{wu98}\looseness=-1
\begin{eqnarray}
\lefteqn{ i\partial_t\mathcal K_0({\bf R},{\bf r},t)\sim -2\Delta_r\mathcal K_0
+8\pi a\rho_0\Phi_0({\bf R})^2 \delta({\bf r})}\nonumber\\
&& + 2[-8\pi a\rho_0\zeta-\zeta_e+V_e({\bf R})+16\pi a\rho_0\Phi_0({\bf R})^2]\mathcal K_0\nonumber\\
&& +8\pi a\rho_0\Phi_0({\bf R})^2\int{\rm d}{\bf w}\, 
\mathcal K_0({\bf R},{\bf w},t)\,\mathcal K_0({\bf R},{\bf w}-{\bf r},t)~.\label{eq:K0-slow}
\end{eqnarray}

The variable ${\bf R}$ enters~(\ref{eq:K0-slow}) as a parameter. By virtue of~(\ref{eq:Phi-app})
and~(\ref{eq:E-zeta}), the term in the brackets of~(\ref{eq:K0-slow}) reads
\begin{equation}
-8\pi a\rho_0\zeta-\zeta_e+V_e({\bf R})+16\pi a\rho_0 \Phi_0({\bf R})^2=8\pi a\rho_0 \Phi_0({\bf R})^2~.
\label{eq:brack}
\end{equation}
Thus, (\ref{eq:K0-slow}) results from~(\ref{eq:K0-com-sph}) of the translationally invariant case
with the replacement\looseness=-1
\begin{equation}
\rho_0\Rightarrow \rho_0 \Phi_0({\bf R})^2~.\label{eq:replace}
\end{equation}
Therefore, the formulation and calculation of section~\ref{sec:trans-inv} remain essentially the same here
when $\Phi_0({\bf R})\neq 0$. The case with $\Phi_0({\bf R})\equiv 0$ leads to a simplified  evaluation
of $K_0$ as shown below.

In the following, we outline the main results in view of~(\ref{eq:replace}).
We solve~(\ref{eq:K0-slow}) with the initial condition
\begin{equation}
\mathcal K_0({\bf R},{\bf r},t=0)=f({\bf r})~,\label{eq:init-slow}
\end{equation}
where $f$ is assumed symmetric, i.e.
$f({\bf r})=f(-{\bf r})$. Thus,\looseness=-1
\begin{equation}
\mathcal K_0({\bf R},{\bf r},t)=\mathcal K_0({\bf R},-{\bf r},t)~.
\label{eq:inv-sym-K}
\end{equation}

By~(\ref{eq:steady-k}),~(\ref{eq:hatK0-soln})--(\ref{eq:p-def}) and~(\ref{eq:K0-slow}), 
the Fourier transform in ${\bf r}$ of $\mathcal K_0$ is found to be
\begin{eqnarray}
\widehat {\mathcal K}_0({\bf R},{\bf k},t)&=&\widehat{g}_0({\bf R},k)-[1-\widehat{g}_0({\bf R},k)^2]\nonumber\\
\mbox{} && \times \frac{p({\bf R},{\bf k})\,e^{-2i\omega({\bf R},k)\,t}}{1-
\widehat{g}_0({\bf R},k)\,p({\bf R},{\bf k})e^{-2i\omega({\bf R},k)\,t}}~,
\label{eq:hatK0-soln-slow}
\end{eqnarray}
where
\begin{equation}
\widehat g_0({\bf R},k):=-\frac{8\pi a\rho_0 \Phi_0({\bf R})^2}{k^2+8\pi a\rho_0\Phi_0({\bf R})^2+
k\sqrt{k^2+16\pi a\rho_0\Phi_0({\bf R})^2}}~,\label{eq:hatg-slow}
\end{equation}
\begin{equation}
\omega({\bf R},k)=k\sqrt{k^2+16\pi a\rho_0 \Phi_0({\bf R})^2}~,\label{eq:omega-slow}
\end{equation}
\begin{equation}
p({\bf R},{\bf k}):=\frac{\widehat g_0({\bf R},k)-\widehat f({\bf k})}{1-\widehat g_0({\bf R},k)\widehat f({\bf k})}~.
\label{eq:p-slow}
\end{equation}
The inversion of~(\ref{eq:hatK0-soln-slow}) is carried out through integral~(\ref{eq:Inv-FT}).

It is worthwhile noting that the $\mathcal K_0$ obtained from~(\ref{eq:hatK0-soln-slow})
is not differentiable in ${\bf R}$ at the boundary $\partial\mathcal R$.
This feature is due to approximation~(\ref{eq:Phi-app}). A remedy is to retain $\Delta_R$
in the equation of motion for the pair-excitation function and describe the spatial changes of $K_0$
along the local normal to $\partial\mathcal R$. This analysis would also require the analogous modification
of the approximation for $\Phi({\bf R})$ described in~\cite{wu98}. These studies lie beyond the scope
of this paper.

With regard to ${\bf R}$, we distinguish the following cases.
\vskip10pt

\noindent{\bf Exterior of region $\mathcal R$} (${\bf R}\notin \mathcal R$).
Formula~(\ref{eq:hatK0-soln-slow}) is greatly simplified since $\Phi_0\equiv 0$. 
We have \looseness=-1
\begin{equation}
\widehat{\mathcal K_0}({\bf R},{\bf k},t)=\widehat f({\bf k})\,e^{-2ik^2\,t},\qquad {\bf R}\notin \mathcal R~.
\label{eq:FTK-slow-out}
\end{equation}
Recall that $\mathcal R$ is defined by~(\ref{eq:regR-def}). By inverting~(\ref{eq:FTK-slow-out})
we obtain the ${\bf R}$-independent function
\begin{equation}
\mathcal K_0({\bf R},{\bf r},t)=\int{\rm d}{\bf r}'\ f({\bf r}')\
\frac{\displaystyle{\exp\biggl(-\frac{|{\bf r}'-{\bf r}|^2}{8t}\biggr)}}{(8\pi t)^{3/2}}~,\qquad 
{\bf R}\notin \mathcal R~,\label{eq:K-slow-out}
\end{equation}
which is the solution of the linear diffusion equation in the three-dimensional 
space~\cite{evans}; cf~(\ref{eq:K0-slow}) with $\Phi_0\equiv 0$. 
This $\mathcal K_0$ vanishes as $t\to +\infty$ for a reasonably wide class of initial data.
In particular, if $f({\bf r})$ is of compact support with size $L$, $t\gg L^{2}$ and $r\ge O(\sqrt{t})$,
then~(\ref{eq:K-slow-out}) yields the known similarity solution
\begin{equation}
\mathcal K_0({\bf R},{\bf r},t)\sim (8\pi t)^{-3/2}\,\exp\biggl(-\frac{r^2}{8t}\biggr)\,
\int {\rm d}{\bf r}'\ f({\bf r}')~.\label{eq:simil}
\end{equation}
If $f\equiv 0$ then $\mathcal K_0$ by~(\ref{eq:K-slow-out}) vanishes identically outside $\mathcal R$.
\vskip10pt

\noindent{\bf Interior of $\mathcal R$} (${\bf R}\in \mathcal R$).
Equation (\ref{eq:hatK0-soln-slow}) is inverted through~(\ref{eq:Inv-FT})
but the integration result is not expressed in simple closed form.
For spherically symmetric initial data, $f({\bf r})=f_0(r)$, $\mathcal K_0({\bf R},{\bf r},t)$ 
is given by~(\ref{eq:hatK0-exp}) replacing $\widehat{\mathcal K}$ by $\widehat{\mathcal K}_0$,
$\omega(k)$ by $\omega({\bf R},k)$, $\widehat g_0(k)$ by $\widehat g_0({\bf R},k)$, and $p_0(k)$ by
\begin{equation}
p_0({\bf R},k)=\frac{\widehat g_0({\bf R},k)-\widehat f_0(k)}{1-\widehat g_0({\bf R},k)\,\widehat f_0(k)}~.
\label{eq:p0-slow}
\end{equation}
For zero initial data, $f_0\equiv 0$, we have $p_0({\bf R},k)=\widehat g_0({\bf R},k)$.

The asymptotic analysis of section~\ref{subsec:asymp} for $f_0\equiv 0$ is applicable here
as well. The non-dimensional variables (scaled distance and time) 
$\tr$ and $\tau$ are now defined by
\begin{equation}
\tr({\bf R}):=[16\pi a\rho_0 \Phi_0({\bf R})^2]^{1/2}r~,\quad \tau({\bf R}):=16\pi a\rho_0\Phi_0({\bf R})^2\,t~.
\label{eq:tau-r-slow}
\end{equation}
These definitions lead again to the integral~(\ref{eq:Lambda-eta}) where $\Lambda$
is now defined by
\begin{equation}
\Lambda({\bf R},r,t):=\big(16\pi a\rho_0\Phi_0({\bf R})^2\big)^{-3/2}[\mathcal K_0({\bf R},{\bf r},t)-
\mathcal K_0({\bf R},{\bf r},+\infty)],\label{eq:Lambda-slow}
\end{equation}
provided that ${\bf R}$ does not lie too close to $\partial\mathcal R$.
The limit $K_0({\bf R},{\bf r},+\infty)$ is the inverse Fourier transform, $g_0({\bf R},r)$,
of $\widehat g_0({\bf R},k)$. Asymptotic formula~(\ref{eq:Lambda-asymp}) applies accordingly.\looseness=-1

\section{Conclusion}
\label{sec:conclusion}

We studied aspects of the time-dependent pair-excitation function, $K_0$, introduced 
by Wu~\cite{wu61,wu98} for interacting Bosons at zero temperature. The particles are trapped 
by a sufficiently slowly varying external potential. By assuming that the wavefunction, 
$\Phi$, of the macroscopic state satisfies a time-independent nonlinear Schr\"odinger equation, 
we determined an approximate solution for the time-dependent integrodifferential equation for $K_0$. Our analysis
relied on the fact that, because of the slowly varying external potential,
the space variables in $K_0$ are separated into the fast variable ${\bf r}={\bf x}-{\bf y}$
and the slow variable ${\bf R}=({\bf x}+{\bf y})/2$ of the center-of-mass system.
The ensuing solution for $K_0$ is given in terms of the Fourier transform in ${\bf r}$
where ${\bf R}$ enters as a vector parameter.\looseness=-1

For zero pair excitation ($K_0=0$) initially, ${\bf R}$ lying inside the trap and 
sufficiently large $t$, i.e. $t\gg \big(16\pi a\rho_0\Phi_0({\bf R})^2\big)^{-1}$
where $\Phi_0({\bf R})$ is given by~(\ref{eq:Phi-app}), we derived 
asymptotic formula~(\ref{eq:Lambda-asymp}). This formula describes how $K_0$ approaches its steady-state values,
$g_0({\bf R},r)$. This result involves a convergent series
containing Lommel's functions with arguments depending on the (scaled) variables $\tr$ and $\tau$ 
of~(\ref{eq:tau-r-slow}). Denoting $[16\pi a\rho_0\Phi_0({\bf R})^2]^{1/2}$ by $u({\bf R})$, 
where $u$ becomes the sound velocity when $\Phi_0$ is unity~\cite{leehuangyang}, 
the asymptotic analysis reveals that $|K_0-g_0|$ takes small, $O(t^{-4})$ values 
for $r\ll 2u\,t$ but increases appreciably to become $O\big((rt)^{-1}\big)$ as $r$ approaches $2ut$. 

The condition $r- n\,2ut=O\big((nt)^{1/3}\big)$,
where $n$ is any positive integer, plays a key role in our asymptotic analysis: it
signifies that a stationary-phase point in one of the requisite integrals 
falls too close to the endpoint of integration.
This effect is reminiscent of the coalescence of critical points
and the onset of caustics in diffraction theory~\cite{berry}, but the special functions
involved are of course different. A physical interpretation of such coalescence in the present case
is tempting but elusive. \looseness=-1
 
Our assumption of $\Phi$ satisfying the nonlinear Schr\"odinger equation is a crucial one.
Clearly, $\Phi$ is modified when $K_0$ acts back on it. This issue is not addressed by
our analysis. Furthermore, the assumption of a time-independent, slowly varying external
potential may have to be relaxed in order to account for a wider class of atomic traps.
This consideration is left for future study.

\section*{Acknowledgements}
I am indebted to Tai T Wu for useful discussions.

\appendix

\section{Solution of differential equation for $\widehat{\mathcal K}$}
\label{app:A}

In this appendix we solve the ordinary differential equation~(\ref{eq:K0-FT-ODE}).

Consider the ordinary differential equation
\begin{equation}
i\,\dot z=z^2+2\big(\alpha^2 +1\big)z+1~,\label{eq:ODE}
\end{equation}
where $z=z(t)$, $\alpha$ is a positive constant and the dot on top of $z$ denotes differentiation with respect to $t$. 
By factorizing the right-hand side of~(\ref{eq:ODE}) we have
\begin{equation}
i\,\dot z= \big( z+1+\alpha^2-\alpha\sqrt{2+\alpha^2}\big)\big(z+1
+\alpha^2+\alpha\sqrt{2+\alpha^2}\big)~.
\label{eq:ODE-fac}
\end{equation}
Separating variables, direct integration of the last equation yields the general solution
\begin{equation}
\frac{i}{2\alpha\sqrt{2+\alpha^2}}\,\ln\biggl|\frac{z+1+\alpha^2-\alpha\sqrt{2+\alpha^2}}
{z+1+\alpha^2+\alpha\sqrt{2+\alpha^2}}\biggr|=t+{\tilde C}~,\label{eq:ODE-sep}
\end{equation}
which is in turn solved for $z(t)$ to give
\begin{equation}
z(t)=\alpha\sqrt{2+\alpha^2}-1-\alpha^2+2\alpha\sqrt{2+\alpha^2}\ 
\frac{Ce^{-i2\alpha\sqrt{2+\alpha^2}\,t}}
{1-Ce^{-i2\alpha\sqrt{2+\alpha^2}\,t}}~,\label{eq:ODE-soln}
\end{equation}
where $C$ is an integration constant. 

The constant $C$ is computed by applying the initial condition $z(0)=z_0$. We find
\begin{equation}
C=\frac{z_0+1+\alpha^2-\alpha\sqrt{1+\alpha^2}}{1+\alpha^2-\alpha\sqrt{2+\alpha^2}}~.
\label{eq:C}
\end{equation}
Equation~(\ref{eq:hatK0-soln}) is recovered by identifying $\alpha$
with $(8\pi a\rho_0)^{-1/2}\,k$ and $z_0$ with $\widehat f({\bf k})$.\looseness=-1

\section{Complex singularities of $\widehat K$ for zero initial data}
\label{app:B}

In this appendix we discuss the analytic continuation of the right-hand side of~(\ref{eq:hatK0-symm})
in the complex $k$-plane. We focus on the simplified yet physically appealing case with $p(k)=\widehat g_0(k)$,
which results from zero initial data, $f({\bf r})\equiv 0$ in~(\ref{eq:init}).

In units where $16\pi a\rho_0=1$ the function of interest reads
\begin{equation}
U(k)=\widehat g_0(k)+4\omega(k)\ \frac{\widehat g_0(k)^2\,
e^{-2i\omega(k)\,t}}{1-\widehat g_0(k)^2\,e^{-2i\omega(k)\,t}}~,\label{eq:U-def}
\end{equation}
where $\omega(k)=k\sqrt{k^2+1}$ and
\begin{equation}
\widehat g_0(k)=-2\big(k^2+\textstyle{\frac{1}{2}}-\omega\big)~,\quad
\widehat g_0(k)^{-1}=-2\big(k^2+\textstyle{\frac{1}{2}}+\omega\big)~.
\label{eq:g0-alt}
\end{equation}
After some straightforward algebra, we obtain the formula
\begin{equation}
U(k)=\frac{-i\,{\mathcal S}(\omega)}{i(k^2+\frac{1}{2})\,{\mathcal S}(\omega)+\cos(\omega t)}~,
\label{eq:U-form}
\end{equation}
where
\begin{equation}
{\mathcal S}(\omega):=\frac{\sin(\omega t)}{2\omega}~.\label{eq:S-def}
\end{equation}
Because ${\mathcal S}(\omega)$ is analytic and even in $\omega$, the only 
possible singularities of $U(k)$ are poles. Note that the branch points $k=0$ and $\pm i$ of $\omega(k)$
are regular points of $U(k)$.

Next, we discuss the location of the poles of $U(k)$.
With the substitution
$k=\sinh\eta~$, we find that the denominator in~(\ref{eq:U-def}) vanishes at points $\eta$ where
\begin{equation}
\textstyle\frac{1}{2} t\,\sinh (2\eta)-i\,2\eta =m\pi,\quad m=\pm 1,\,\pm
2,\,\ldots\,~.\label{eq:poles-I}
\end{equation}
Let $\eta=a+ib$. The imaginary part of the last equation yields
\begin{equation}
4a=t\,\sin (2b)\,\cosh (2a)\qquad (t>0)~.\label{eq:ReIm}
\end{equation}
It suffices to consider only the range $-\pi/2< b \le \pi/2$.
It is readily concluded that it is impossible to have $a>0$ and 
$-\pi/2 < b < 0$; or, $a<0$ and $0< b < \pi/2$. Thus, the poles must lie
in the first and third quadrants of the complex $k$-plane. A further examination
of~(\ref{eq:poles-I}) shows that these poles are simple.

It is of interest to describe the poles analytically for
small and large values of $|m\pi|/t$. For $t\gg 1$ and $-\pi< 2b<\pi$, we find that~(\ref{eq:poles-I}) 
is solved approximately by
\begin{eqnarray}
\eta &\sim &  \frac{m\pi}{t-2i},\quad |m|\pi\ll t,\quad |m|= 1,\,2\,\ldots\,~,\nonumber\\
\eta &\sim &  \frac{1}{2}\,{\rm sg}(m)\,\left( 1+\frac{i}{|m|\pi}\right)\,\ln
\frac{4|m|\pi}{t},\quad |m|\pi\gg t~,\label{eq:poles-II}
\end{eqnarray}
where ${\rm sg}$ is the sign function, i.e. ${\rm sg}(x)=1$ if $x>0$ and ${\rm sgn}(x)=-1$ if $x<0$. 
Evidently, the $k=\sinh\eta$ corresponding to~(\ref{eq:poles-II}) lies in the first or
third quadrant of the $k$ plane. For fixed $t$, these poles approach the real axis.

\section{Analytic continuation formulas for $S_{0,\frac{1}{3}}(z)$}
\label{app:C}

In this appendix, $\,S_{0,\frac{1}{3}}(z\,e^{-i\pi})$ and
$\,S_{0,\frac{1}{3}}(z\,e^{i2\pi})$, which are involved in the asymptotics for
$\Lambda(r,t)$ in section~\ref{subsec:asymp}, are expressed in terms of
$S_{0,\frac{1}{3}}(z)$. 

The starting point is the relation \cite{bateman}
\begin{eqnarray}
S_{0,\frac{1}{3}}(z)  &=&   s_{0,\frac{1}{3}}(z)+{\textstyle\frac{1}{2}}
\Gamma({\textstyle\frac{1}{2}}-{\textstyle\frac{1}{6}})\,
\Gamma({\textstyle\frac{1}{2}}+{\textstyle\frac{1}{6}})\,
[-\sin(\pi/6)\,J_{1/3}(z)\nonumber\\
&&\mbox{}\quad -\cos(\pi/6)\,Y_{1/3}(z)]\nonumber\\
{}  &=&   s_{0,\frac{1}{3}}(z)-\frac{\pi}{2}\,[\tan(\pi/6)\,
J_{1/3}(z)+Y_{1/3}(z)]~,\label{eq:Lomm-reln}
\end{eqnarray}
where $\Gamma(z):=\int_0^{\infty}{\rm d}x\,x^{z-1}e^{-x}$ is the Gamma function,
$J_\nu(z)$ and $Y_{\nu}(z)$ are Bessel functions~\cite{bateman},
\begin{equation}
s_{0,\frac{1}{3}}(z):=\frac{9z}{8}\ {}_1F_2(1;4/3,5/3;-z^2/4)~,\label{eq:sm-s-def}
\end{equation}
and ${}_1F_2$ is a hypergeometric series~\cite{batemanI}. Evidently, we have
\begin{equation}
s_{0,\frac{1}{3}}(z\,e^{-i\pi})=-s_{0,\frac{1}{3}}(z)=-s_{0,\frac{1}{3}}(z\,
e^{i2\pi})~.
\label{eq:sm-s-prop}
\end{equation}
Useful analytic continuation formulas for the Bessel functions are~\cite{bateman}
\begin{eqnarray}
J_{1/3}(z\,e^{-i\pi})&=&e^{-i\pi/3}\,J_{1/3}(z)~,\quad
J_{1/3}(z\,e^{i2\pi})=e^{i2\pi/3}\,J_{1/3}(z)~,\nonumber\\
Y_{1/3}(z\,e^{-i\pi})  &=&   e^{i\pi/3}\,Y_{1/3}(z)-i\,J_{1/3}(z)~,\nonumber\\
Y_{1/3}(z\,e^{i2\pi})  &=&   e^{-i2\pi/3}\,Y_{1/3}(z)
+i\,J_{1/3}(z)~.\label{eq:Bessel-analt}
\end{eqnarray}

By combining~(\ref{eq:Lomm-reln})--(\ref{eq:Bessel-analt}) we obtain the desired analytic
continuation formulas,
\begin{equation}
S_{0,\frac{1}{3}}(z\,e^{-i\pi})=-S_{0,\frac{1}{3}}(z)-\frac{\pi}{2}\,\sqrt{3}
\,e^{-i\pi/3}\, H_{1/3}^{(1)}(z)~,\label{eq:Hankel-I}
\end{equation}
\begin{equation}
S_{0,\frac{1}{3}}(z\,e^{i2\pi})=S_{0,\frac{1}{3}}(z)+\frac{\pi}{2}\,\sqrt{3}\,
e^{-i\pi/3}\,H_{1/3}^{(1)}(z)~,\label{eq:Hankel-II} 
\end{equation}
where $H_{\nu}^{(1)}(z)$ is the Bessel function of the third kind~\cite{bateman}. 

Equations~(\ref{eq:Hankel-I}) and~(\ref{eq:Hankel-II}) hold for any $z$. Replacing $z$
by $i\,z$ we have
\begin{eqnarray}
S_{0,\frac{1}{3}}(i\,z\,e^{-i\pi})&=& -S_{0,\frac{1}{3}}(i\,z)+ \sqrt{3}\,
K_{1/3}(z)~,\label{eq:mod-HankI}\\
S_{0,\frac{1}{3}}(i\,z\,e^{i2\pi})&=& S_{0,\frac{1}{3}}(i\,
z)-\sqrt{3}\,K_{1/3}(z)~,\label{eq:mod-Hank}
\end{eqnarray}
where $K_{\nu}(z)$ is the modified Bessel function of the third kind~\cite{bateman}.

\end{document}